\documentclass[prb,amsmath,superscriptaddress,twocolumn,showpacs]{revtex4}
\usepackage[dvips]{graphicx} 
\usepackage{bm}
\usepackage{pifont}
\usepackage{amssymb}
\usepackage{colordvi}
\usepackage{graphicx}
\usepackage{color}

\definecolor{orange}{rgb}{1,0.5,0}
\definecolor{darkgreen}{rgb}{0,0.5,0}
\definecolor{darkblue}{rgb}{0,0,0.5}
\definecolor{purple}{rgb}{0.35,0,0.35}

\usepackage{hyperref}

\begin{document}

\title{Full-counting statistics for molecular junctions: \\
the fluctuation theorem and singularities \\
}

\author{Y. Utsumi}
\email{utsumi@phen.mie-u.ac.jp }
\affiliation{Department of Physics Engineering, Faculty of Engineering, Mie University, Tsu, Mie, 514-8507, Japan
}
\affiliation{Kavli Institute for Theoretical Physics China, CAS, Beijing 100190, China}

\author{O. Entin-Wohlman}
\affiliation{
Physics Department, Ben Gurion University, Beer Sheva 84105, Israel
}
\affiliation{
Raymond and Beverly Sackler School of Physics and Astronomy, Tel Aviv University, Tel Aviv 69978, Israel
}
\affiliation{Kavli Institute for Theoretical Physics China, CAS, Beijing 100190, China}

\author{A. Ueda}
\affiliation{Faculty of Pure and Applied Sciences, Division of Applied Physics,
University of Tsukuba, Tsukuba, Ibaraki, 305-8573, Japan}

\author{A. Aharony}
\affiliation{
Physics Department, Ben Gurion University, Beer Sheva 84105, Israel
}
\affiliation{
Raymond and Beverly Sackler School of Physics and Astronomy, Tel Aviv University, Tel Aviv 69978, Israel
}
\affiliation{Kavli Institute for Theoretical Physics China, CAS, Beijing 100190, China}

\date{\today}

\begin{abstract}
We study the full-counting statistics of charges transmitted through a single-level quantum dot weakly coupled to a local Einstein phonon which causes fluctuations in the dot energy. 
An analytic expression for the cumulant generating-function, accurate up to  second order in the electron-phonon coupling and valid for finite voltages and temperatures, is obtained in the extended wide-band limit. 
The result accounts for nonequilibrium phonon distributions induced by the source-drain bias voltage, and concomitantly satisfies the fluctuation theorem. 
Extending  the counting field to the complex plane, we investigate the locations of possible singularities of the cumulant generating-function, and exploit them to identify regimes in which the electron transfer is affected differently by the coupling to the phonons. 
Within a large-deviation analysis, we find a kink in the probability distribution, analogous to a first-order phase transition in thermodynamics, which would be a unique hallmark of the electron-phonon correlations. 
This kink reflects the fact that although inelastic scattering by the phonons once the voltage exceeds their frequency can scatter electrons opposite to the bias, this will never generate current flowing against the bias at zero temperature, in accordance with the fluctuation theorem.

\end{abstract}

\pacs{05.30.-d, 72.70.+m, 71.38.-k, 73.63.Kv, 05.70.Ln}

\maketitle

\section{Introduction} 
\label{INTRO}


It has been revealed in electric transport measurements that the electron-phonon interaction induces unique  features in the nonequilibrium current through molecular junctions~\cite{Smit,Kiguchi} and atomic wires.~\cite{Tal,Agrait,Kumar} 
In particular, when the phonon energy is small compared to the resonance width on the junction, the inelastic phonon scattering increases/decreases the current for small/large transmission probabilities as the source-drain bias voltage $V$ exceeds the local phonon frequency $\omega_0$.~\cite{Smit,Tal,Agrait,Kumar}
Quite a number of theoretical microscopic models, e.g. Ref.~\onlinecite{Viljas,delaVega}, including density-functional theories,~\cite{Frederiksen} have been devoted to such junctions (see Ref.~\onlinecite{Galperin} and references therein). 
They revealed that a simplified model,~\cite{Egger} of a single-level quantum dot coupled to a local Einstein phonon mode causing fluctuations of the dot energy level,~\cite{Holstein} seems to suffice to capture this ubiquitous feature. 
Recently, the current noise of an atomic wire has been measured,~\cite{Kumar} and it was observed that the electron-phonon interaction can enhance or reduce the noise, depending on the value of the transmission probability. 
Based on the theory of Avriller and Levy Yeyati,~\cite{Avriller} the negative correction is understood as resulting from the anti-bunching of two electrons: 
An electron cannot be inelastically scattered by a phonon when the final state is already occupied by another electron. 


Avriller and Levy Yeyati considerations follow from the theory of full-counting statistics (FCS),~\cite{Levitov,Nazarov} which is most convenient for analyzing  nonequilibrium electric transport. 
Indeed, considerable effort has been invested in recent years in exploiting FCS to study various aspects of nonequilibrium quantum transport [e.g. Refs.~\onlinecite{Avriller,Levitov,Nazarov,Komnik,Gogolin,BUGS,Bagrets,UGS,Sakano,Haupt1,Haupt2,Novotny,Schmidt,Urban,Simine,Schaller,Maier} and references therein]. 
FCS refers to the probability distribution $P_\tau(q)$ of the charge $q$ to be transmitted through a quantum conductor during a certain measurement time $\tau$ at out-of-equilibrium conditions (we set $e=\hbar =1$). 
The effect of coupling to a vibrational mode on the electric transport has also been investigated  in this context, both for a weak~\cite{Haupt1,Haupt2,Novotny,Avriller,Schmidt,Urban} and  a strong~\cite{Simine,Schaller,Maier} electron-phonon coupling.


In general, it is a rather formidable task to calculate the  FCS of interacting electrons. 
For this reason, most of the investigators have taken advantage of the Keldysh field-theory technique. 
There, the characteristic function (CF), 
\begin{align}
{\cal Z}^{}_{\tau}(\lambda )=\sum_{q}P^{}_{\tau}(q)e^{iq\lambda}
\ , 
\label{Z}
\end{align}
or the scaled cumulant generating-function (CGF)~\cite{Touchette} pertaining to the steady state, 
\begin{align}
{\cal F}(\lambda)
=
\lim_{\tau \rightarrow \infty}
\frac{1}{\tau} \ln{\cal Z}^{}_{\tau}(\lambda) \ , 
\label{F}
\end{align}
can be formally written as a `partition function' or a `free energy', respectively,  defined along the Keldysh time-contour. 
The  $\lambda$ appearing in these formulae is termed  the counting field or the counting parameter. 
However, even if one calculates the CGF successfully, one still needs to find a way to characterize the electronic correlations in it. 
One promising approach would be to utilize the location distribution of the zeros of the CF, or equivalently that of the singularities of the CGF, by allowing the counting field $\lambda$ in Eq.~(\ref{Z}) to attain complex values,~\cite{Abanov,Kambly,DIvanov,FG}
similarly to the Yang-Lee theory of phase transitions.~\cite{Yang}
This idea is based on the recent observation that, upon transforming $\lambda$ into $u$, 
\begin{align}
u={\rm e}^{i \lambda} , 
\label{cft}
\end{align}
the singularities of the CGF of noninteracting electrons transported between two terminals are all on the negative real axis of the $u-$plane.~\cite{Abanov,Vanevic,Hassler} 
It suggests that singularities off the negative real axis would characterize electronic correlations. 
From this aspect, molecular junctions are rather advantageous since second-order perturbation theory in the electron-phonon coupling would capture most relevant features of the electron-phonon correlations in them, allowing for obtaining the location distribution of the singularities. 


Another recent ingredient is the fluctuation theorem (FT).~\cite{EG,Tobiska,Foerster,SU,US,Andrieux,Esposito,Campisi,Altland,Lopez,UK,Nakamura,Saira} 
The FT is a consequence of  micro-reversibility and  can be understood as a microscopic extension of the second law of thermodynamics. 
Despite its simple appearance, a detailed-balance like relation~[see Eq.~(\ref{FT}) below], 
the FT reproduces the linear-response results, i.e. it ensures the fluctuation-dissipation theorem and Onsager's reciprocal relations close to equilibrium,~\cite{Tobiska,Foerster,SU,Andrieux,Esposito,Campisi} while conveying invaluable information at nonequilibrium conditions. 
For molecular junctions, the FT has been addressed using the master-equation approach of FCS for incoherent electron transport.~\cite{Schaller,Simine}
The FT is considered to be a basic symmetry, such as  gauge invariance, which the CF should fulfill.


In the present paper we investigate the FCS of electrons coupled to phonons under out-of-equilibrium conditions. 
The quantity to be calculated and analyzed is the cumulant generating-function.~\cite{Haupt1,Haupt2,Novotny,Avriller,Schmidt,Urban} 
Employing the Luttinger-Ward functional, \cite{Baym,Luttinger,Ivanov}, we obtain its full analytic expression, or equivalently {\it all} cumulants, in the wide-band limit, treating systematically the nonequilibrium phonon distribution via a self-consistent condition accurate up to second order in the electron-phonon coupling.
There are other attempts in the literature \cite{Novotny,Urban} to account for the nonequilibrium phonon effect avoiding the Luttinger-Ward functional;  these 
produce contradicting results; our first two cumulants agree with those of Ref. \onlinecite{Novotny}, which has not gone beyond the second cumulant. 
(Our results disagree with those of Ref. \onlinecite{Urban}.)

The structure of the paper is as follows.
We begin in Sec. \ref{SECFT} with brief general explanations of the FT, the large-deviation analysis and singularities of the CGF. 
We then introduce in Sec. \ref{CGF} the model Hamiltonian and present analytical results for the CGF and detailed explanations of the calculations and the  approximations involved. 
A self-consistent calculation based on the nonequilibrium Luttinger-Ward functional,~\cite{Baym,Ivanov,Luttinger} is relegated to Appendices~\ref{LW} and \ref{LCE}. 
In Sec. \ref{RESU} we analyze the singularities of the CGF and demonstrate the probability distribution within the large-deviation analysis. 
Section~\ref{SUMM} summarizes our results. Technical details of the calculations are given in Appendices \ref{CEP} and \ref{PHPR}.

\section{The fluctuation theorem and singularities}
\label{SECFT}

The definition of the probability distribution $P_{\tau}(q)$ in quantum systems requires special care.~\cite{Levitov,Nazarov} 
Full-counting statistics theory begins with by introducing  the CF [see Eq. (\ref{Z}) and  Eq.~(\ref{ZKE}) below for the definition employed in the Keldysh technique] and then defining the quasi-probability distribution by the inverse Fourier transform of the CF, 
\begin{align}
P^{}_{\tau}(q)
=
\frac{1}{2 \pi} \int_{-\pi}^\pi d \lambda 
{\rm e}^{-i \lambda q} {\mathcal Z}_\tau(\lambda)
\ .
\label{ifotr}
\end{align}
More details are given in Sec. \ref{CGF}. 
The $n$-th cumulant, in  steady state,  is given by the $n$-th derivative of the CGF, Eq.~(\ref{F}), 
\begin{align}
\langle\!\langle I^n \rangle\!\rangle 
=
\lim_{\tau \rightarrow \infty} 
\frac{\langle\!\langle q^n \rangle\!\rangle}{\tau}
=
\frac{\partial^n {\cal F}(\lambda )}{\partial (i\lambda )^n }
\Big |^{}_{\lambda =0}\ .
\end{align}
For example, the average current is  the first cumulant, and the current noise is the second one. 

The FT in the context of quantum electric transport relates the probability of the current  to flow oppositely to the bias voltage because of thermal agitations, $P_{\tau}(-q)$, \cite{Tobiska,Foerster,Andrieux,Esposito,Campisi,SU,Altland,Lopez,UK,Nakamura,US} to the distribution $P_{\tau}(q)$, 
\begin{align}
P^{}_{\tau}(-q)=P^{}_{\tau}(q)e^{-\beta q V}\ ,
\label{FT} 
\end{align}
where $\beta $ is the inverse temperature.  The FT can be equivalently written in terms of the CGF ${\cal F}$, Eq. (\ref{F}), as 
\begin{align}
{\cal F}(\lambda )={\cal F}(-\lambda +i\beta V) \, . 
\label{FFT}
\end{align}

\begin{figure}[ht]
\includegraphics[width=.9 \columnwidth]{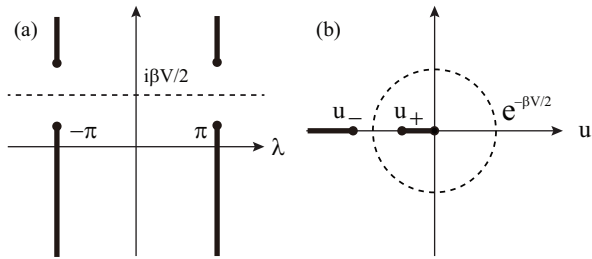}
\caption{
The thick
lines represent branch cuts corresponding to the continuous singularities of the CGF of noninteracting electrons, Eq.~(\ref{ZO}), in the $\lambda-$plane (a) and in the $u-$plane (b). 
The FT ensures that the branch cuts are symmetrically distributed around $\lambda = i \beta V/2$ in the  $\lambda-$plane (a). 
In the $u-$plane (b) the FT relates the  branch cuts located  inside and outside the dotted circle. 
}
\label{uplane1}
\end{figure}

The relation (\ref{FFT}) restricts the possible locations of the singularities of the CGF in the $\lambda-$plane. 
As an example, we depict in Fig.~\ref{uplane1} (a) the branch cuts corresponding to the continuous singularities of the CGF pertaining to two-terminal transport of noninteracting electrons, Eq.~(\ref{ZO}) below. 
This CGF is $2 \pi-$periodic along the real axis of $\lambda$, which guarantees integer values of charge.~\cite{Levitov} 
The branch cuts, depicted by thick lines,  are at  ${\rm Re} \, \lambda = (2 n-1) \pi$, where $n$ is an integer. 
The FT ensures that the branch cuts are symmetrically distributed around $\lambda = i \beta V/2$. 
(In Fig.~\ref{uplane1} (a), the upper left thick line is identical to the lower right thick line, etc.) 
The $2 \pi-$periodicity is removed by the conformal transformation Eq.~(\ref{cft}). 
Then the branch cuts are on the negative real axis of the $u$-plane [Fig.~\ref{uplane1} (b)].~\cite{Vanevic,Abanov,Kambly,Hassler}

The steady-state probability distribution, beyond the central-limit theorem, is derived within the theory of large deviations.~\cite{Touchette}
At steady state, realized in the $\tau \to \infty$   limit, 
we scale $q=I \tau$ and ${\mathcal Z}_\tau \approx {\rm e}^{\tau {\mathcal F}}$. 
Then the integral of the inverse Fourier transform Eq.~(\ref{ifotr}) can be estimated by the saddle-point approximation~\cite{BO} and the result is written with the rate function~\cite{Touchette} ${\mathcal I}$ as $P^{}_{\tau} \approx {\rm e}^{-\tau {\mathcal I}}$. 
Since $P_\tau$ is real and positive, the saddle point is expected to reside on the imaginary axis of complex $\lambda-$plane. 
Then the rate function is written in the form of a Legendre-Fenchel transform, 
\begin{align}
{\mathcal I}(I)
=
-\lim_{\tau \to \infty} \frac{1}{\tau} \ln P^{}_{\tau}(I \tau)
=
\max_{\lambda} \left \{ i \lambda I -{\cal F}(\lambda) \right \}
\, . 
\label{ratefun}
\end{align}
Here $\lambda$ is a purely imaginary number. 
In most cases, 
the CGF is real, i.e. the imaginary part of the exponent of the Fourier integral (\ref{ifotr}), 
$\ln {\mathcal Z}_\tau(\lambda)-i \lambda q$, 
is zero on the imaginary axis of $\lambda-$plane. 
Then the imaginary axis is expected to be the steepest contour of the integral (\ref{ifotr}).~\cite{BO} 
There are few exceptions where singularities of the CGF are on the imaginary axis~\cite{BUGS,Bagrets} as we will also find below. 
The relation between the CGF and the rate function is analogous to that of thermodynamic potentials. 
It suggests that singularities on the imaginary axis of the $\lambda-$plane would also cause characteristic features in the rate function.

\section{Model Hamiltonian and  cumulant-generating function}
\label{CGF}

\begin{figure}[ht]
\includegraphics[width=.45 \columnwidth]{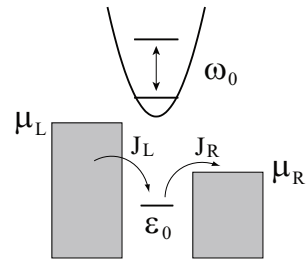}
\caption{Schematic description of the single-level quantum dot coupled to an Einstein vibrational mode.
}
\label{setup}
\end{figure}

Our explicit calculations are carried out for a simple model [Fig.~\ref{setup}], a single-level quantum dot coupled to a local Einstein phonon, which induces fluctuations in  that level energy.~\cite{Mitra,Egger,Holstein,Galperin} 
The model Hamiltonian is
\begin{align}
{\cal H}={\cal H}^{}_{\rm mol}+{\cal H}^{}_{\rm ph}+{\cal H}^{}_{\rm lead}+{\cal H}^{}_{{\rm tun}, \pm}\ , \label{HAM}
\end{align}
where the $\pm$ subscript is due to the presence of counting fields (see below).
The molecular bridge or the atomic wire is described by the Hamiltonian ${\cal H}_{\rm mol}$, 
\begin{align}
{\cal H}^{}_{\rm mol}=[\epsilon^{}_{0}+\gamma (b+b^{\dagger})]c^{\dagger}_{0}c^{}_{0}\ ,
\end{align}
in which $c_{0}$ ($c^{\dagger}_{0}$) destroys (creates) an electron on the localized level representing the molecule, of energy $\epsilon_{0}$, $b$ and $b^{\dagger}$ are the destruction and creation operators of the vibrations to which the electron is coupled while residing on the dot, and $\gamma$ is the strength of the electron-phonon coupling. 
The vibrational modes obey the harmonic Hamiltonian 
\begin{align}
{\cal H}^{}_{\rm ph}=\omega^{}_{0}b^{\dagger}b\ ,
\end{align}
where $\omega_{0}$ is the frequency of the Einstein phonon. 
The leads are represented by free electron gases, of creation and destruction operators $c_{{\rm r}k}^{\dagger}$ and $c_{{\rm r}k}^{}$, and eigen energies $\epsilon^{}_{{\rm r}k}$, 
\begin{align}
{\cal H}^{}_{\rm lead}=\sum_{\rm r=L,R}\sum_{k}\epsilon^{}_{{\rm r}k}c^{\dagger}_{{\rm r}k}c^{}_{{\rm r}k}\ .
\end{align}
Finally there is the tunneling Hamiltonian, coupling the leads to the molecule. 
This part of the Hamiltonian is augmented by the counting fields. 
Those  appear as phase factors on the operators $c^{}_{{\rm r}k}$  and $c^{\dagger}_{{\rm r}k}$, 
\begin{align}
{\cal H}^{}_{{\rm tun, }\pm}=\sum_{\rm r=L,R}\sum_{k}J^{}_{\rm r}e^{\pm i \lambda^{}_{\rm r}}c^{\dagger}_{{\rm r}k}c^{}_{0}+{\rm H.c.}\ ,\label{HTUN}
\end{align}
where $J_{\rm L}$ and $J_{\rm R}$ are the tunneling amplitudes between the left and the right lead, and the dot. 

\begin{figure}[ht]
\includegraphics[width=.45 \columnwidth]{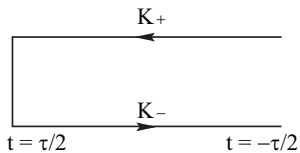}
\caption{The Keldysh contour.
}
\label{Keldysh}
\end{figure}

In terms of the tunneling Hamiltonian Eq.~(\ref{HTUN}), the characteristic function is~\cite{Nazarov} 
\begin{align}
{\cal Z}_\tau(\lambda )
=
\Big \langle T^{}_{\rm K} 
\exp \Bigl (-i\int_{\rm K} dt
{\cal H}_{{\rm tun},\pm}(t)_{\rm I}\Bigr )
\Big \rangle \ ,
\label{ZKE}
\end{align}
where K denotes the Keldysh contour, which runs from $t=-\tau /2$ to $t=\tau /2$ on the upper branch and returns to $t=-\tau/2$ on the lower one  (see Fig. \ref{Keldysh})  and $T_{\rm K}$ is the time-ordering operator along that contour. The subscript I indicates time dependence in the interaction picture.
The $\pm$ notation indicates the  branch of the Keldysh contour on which the tunneling Hamiltonian is effective, $+$ for the upper branch in Fig. \ref{Keldysh} and $-$ for the lower one. 
In the long-time limit, the scaled CGF  depends only  on the the difference of the two counting fields 
\begin{align}
\lambda=\lambda^{}_{\rm L}-\lambda^{}_{\rm R} \ .
\label{LAM}
\end{align}
As $\lambda_{\rm L}+\lambda_{\rm R}$ counts the number of electrons flowing into the dot, the fact that the CGF depends solely on $\lambda$ implies current conservation. 

Since in our case the electron-phonon coupling $\gamma$ is weak, we carry out the calculation up to second-order accuracy, ${\mathcal O}(\gamma^2)$. 
There is a subtle point in this expansion.
A naive second-order perturbation theory is not capable of producing the correct nonequilibrium phonon distribution.~\cite{Frederiksen,Viljas,Haupt1,Haupt2,Novotny,ora,Rosch} 
We therefore have to perform a re-summation of infinite diagrams by adopting the linked cluster expansion, see e.g. Ref. \onlinecite{UGS}, or a more advanced method, the nonequilibrium Luttinger-Ward functional, $\Phi$.~\cite{Baym,Ivanov,Luttinger,Park} 
The first approach had been adopted in Refs. \onlinecite{Schmidt} and \onlinecite{Urban}, while the second had been employed implicitly by Gogolin and Komnik, ~\cite{Gogolin} and hence by later studies~\cite{Haupt1,Haupt2,Novotny,Avriller} based on Ref. \onlinecite{Gogolin}. 
Although the second scheme is physically transparent since it relies on the self energy of the  electron Green function, the price to pay is that a self-consistent calculation is required in order to ensure various conservation laws. 
Here we follow the second approach based explicitly on the Luttinger-Ward functional, which enables us to perform the self-consistent calculations of the CGF in a transparent manner. 
Details, are given in Appendices~\ref{LW} and \ref{LCE}.

In terms of the Luttinger-Ward potential, the cumulant generating-function is
\begin{align}
{\cal F}(\lambda )={\cal F}^{}_{0}(\lambda )-\bar{\Phi}^{(2)}(\lambda)
\ ,
\label{FFF}
\end{align}
where 
 ${\cal F}_{0}$ is the CGF pertaining to  noninteracting electrons, and  given in terms of the Keldysh Green function $g_{\lambda}$ [see Eq. (\ref{INVG})] of the {\em electronic} part of the Hamiltonian Eq. (\ref{HAM}), 
\begin{align}
{\mathcal F}_0(\lambda)
&= 
\frac{1}{2 \pi}
\int d \omega
\ln 
\frac{{\rm det} g_\lambda^{-1}(\omega)}
{{\rm det} g_0^{-1}(\omega)}
\ , 
\label{ZORD1}
\end{align}
with  the trace being performed over the $2 \times 2$ Keldysh space. 
The CGF satisfies the normalization condition 
${\mathcal F}_0(0)=0$. 
The effect of the electron-phonon interaction is included in the scaled Luttinger-Ward potential 
$\bar{\Phi}^{(2)}$, 
which consists of linked diagrams up to 
${\cal O}(\gamma^{2})$ accuracy, 
\begin{align}
\lim_{\tau \to \infty}
\frac{1}{\tau}
{\Phi}(g_\lambda)
=
\bar{\Phi}(\lambda)
=
\bar{\Phi}^{(2)}(\lambda)
+
{\mathcal O}(\gamma^4)
\  . 
\label{slw}
\end{align}

\subsection{The electronic part}

The inverse  of the Keldysh Green function $g_{\lambda}$ reads 
\begin{widetext}
\begin{align}
g^{}_{\lambda}(\omega )^{-1}_{}=\left[\begin{array}{cc}\omega-\epsilon^{}_{0}+i\sum_{\rm r=L,R}\Gamma^{}_{\rm r}[f^{-}_{\rm r}(\omega )-f^{+}_{\rm r}(\omega )]/2&i\sum_{\rm r=L,R}\Gamma^{}_{\rm r}f^{+}_{\rm r}(\omega )e^{i\lambda^{}_{\rm r}}\\
-i\sum_{\rm r=L,R}\Gamma^{}_{\rm r}f^{-}_{\rm r }(\omega )e^{-i\lambda^{}_{\rm r}}&-\omega+\epsilon^{}_{0}+i\sum_{\rm r=L,R}\Gamma^{}_{\rm r}[f^{-}_{\rm r}(\omega )-f^{+}_{\rm r}(\omega )]/2\end{array}\right ]\ ,
\label{INVG}
\end{align}
\end{widetext}
where $\Gamma_{\rm L}$ and $\Gamma_{\rm R}$ are the partial widths of the localized level induced by the coupling with the leads, 
\begin{align}
\Gamma^{}_{r}=2\pi\nu^{}_{\rm r}|J^{}_{\rm r}|^{2}\  ,\ \ \ r={\rm L,R}\ .
\end{align}
Here $\nu_{\rm L}$ and $\nu_{\rm R}$ are the densities of states in the left and right leads. 
Each of the leads is specified by its chemical potential $\mu_{\rm r}$, such that $\mu_{\rm L}-\mu^{}_{\rm R}=V$, and their electron/hole distribution is accordingly given by
\begin{align}
f^{\pm}_{\rm r}(\omega )=\frac{1}{e^{\pm\beta (\omega-\mu^{}_{\rm r})}+1}\ .\label{FERMI}
\end{align}

Our calculation is confined to the extended wide-band limit, \cite{Haupt1,Haupt2,Novotny} 
in which either  the level broadening $\Gamma=\sum_{\rm r=L,R}\Gamma^{}_{\rm r}$ is larger than the all other energy scales except for the dot  level $|\epsilon_0|$ or the dot level $|\epsilon_0|$ is larger than all other energy scales, 
i. e. 
\begin{align}
|V|, k_{\rm B} T, \omega_0 \ll \Gamma \, ,
\label{wbl1}
\end{align}
or 
\begin{align}
|V|, k_{\rm B} T, \omega_0,\Gamma \ll |\epsilon_0|\ .
\label{wbl2}
\end{align}
In both cases, we can neglect the energy dependence of the dot density of states. 
The normalized dot density of states is then replaced by its value  at the Fermi energy [see Eq.~(\ref{DOS})] 
\begin{align}
\rho_0
=
\frac{\Gamma^2}{4 \epsilon_0^2+\Gamma^2}\ 
. 
\label{APPRDOS}
\end{align}
The calculation detailed in Appendix \ref{CEP} yields ~\cite{Levitov} 
\begin{align}
{\cal F}^{}_{0}(\lambda )=
\frac{
({\rm arccosh}^{}X_\lambda)^{2}
}
{2 \pi \beta}
-
\frac{\beta V^2}
{8 \pi}
\ , 
\label{ZO}
\end{align}
where
\begin{align}
X_\lambda
=&
({\mathcal T}-1)
\cosh \frac{\beta V}{2}
-
{\mathcal T}
\cosh \frac{\beta V+2 i \lambda}{2}
\, . 
\label{X}
\end{align}
Here ${\cal T}$ is  the transmission of the localized level,  
\begin{align}
{\cal T}
=
\alpha \rho_0
\ , \label{T_}
\end{align}
written in terms of the normalized density of states Eq.~(\ref{APPRDOS}) and the transmission probability at resonance 
\begin{align}
\alpha &=
\frac{4 \Gamma^{}_{\rm L} \Gamma^{}_{\rm R}}{\Gamma^{2}}
\, . 
\label{AL}
\end{align}
The counting field dependence is only through $X_\lambda$. 
This result obeys the FT since when $\lambda \rightarrow -\lambda +i\beta V$, $X_\lambda$ is unchanged.

This CGF possesses continuous lines of singularities for  $X_\lambda \in [1,\infty)$. 
As discussed in Sec. \ref{SECFT}, in the $u-$plane these become branch cuts $(-\infty,u_-]$ and $[u_+,0)$, 
where the branch points are at 
$u_\pm= 
{\rm e}^{-\beta V/2}(x \pm \sqrt{x^2-1})
$, 
with 
$x=\cosh(\beta V/2) (1-1/{\mathcal T})-1/{\mathcal T}$
[see Figs.~\ref{uplane1} (a) and (b)]. 
The zeroth-order  CGF (\ref{ZO}) takes a particular simple form  for a symmetric bridge $\Gamma_{\rm L}=\Gamma_{\rm R}$ at resonance $\epsilon_0=0$, for which the transmission is perfect ${\mathcal T}=1$ [see Eqs. (\ref{T_}), (\ref{AL}), and (\ref{APPRDOS})]
\begin{align}
{\cal F}^{}_{0}(\lambda )
=
i\lambda (V+i\lambda /\beta )/(2\pi) \ ,
\label{gaussian}
\end{align}
which describes Gaussian thermal fluctuations. 
On the other hand, when the bridge is extremely askew, 
$\Gamma_{\rm R}\ll\Gamma_{\rm L}$ 
or 
$|\epsilon_0| \gg \Gamma$ 
and thus ${\cal T}\ll1$ one may expand the CGF 
to obtain
\begin{align}
{\cal F}^{}_{0}(\lambda )
\approx
\frac{{\cal T}V}{2\pi}
\Bigl (\frac{e^{i\lambda}-1}{1-e^{-\beta V}}+
\frac{e^{-i\lambda}-1}{e^{\beta V}-1}
\Bigr ) \ , 
\label{bidpoisson}
\end{align}
which is  the bi-directional Poisson form.

\subsection{The phonon-induced part}

\begin{figure}[hb]
\includegraphics[width=.5 \columnwidth]{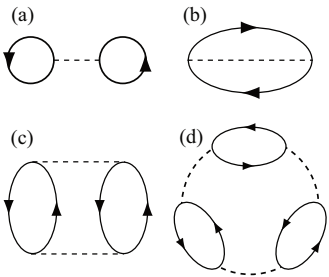}
\caption{Second-order diagrams, the Hartree (a) and Fock (b) terms. 
Diagram (c) is one of the fourth-order ones, and diagram
(d) is one of sixth-order diagrams (a ring diagram). 
The full lines denote electron propagators, and the dashed ones phonon propagators.
}
\label{diagrams}
\end{figure}

The nonequilibrium Luttinger-Ward functional ${\Phi}$ in Eq. (\ref{slw}) results from the coupling of the charge carriers to the vibrational modes. We expand it diagrammatically in powers of the small parameter (see Appendix \ref{LCE} for details)
\begin{align}
g =
\frac{2 \gamma^2}{\pi \Gamma^2}
\, . \label{PARAG}
\end{align}
The relevant diagrams are shown in Fig. 
~\ref{diagrams}:  the second-order diagrams [(a) and (b)]
and two of higher-order diagrams [(c) and (d)]. 
Diagram (a) represents the Hartree term, which is ignored below since it is independent of the phonon distribution [see Eq.~(\ref{hartree})].
Diagram (b) represents the Fock term, which depends on the phonon distribution function at equilibrium [see Eq.~(\ref{fock})]. 
The actual nonequilibrium phonon distribution function can be obtained only by summing up to   an infinite  order (in the electron-phonon coupling) of diagrams. 
The simplest way is to collect all ring-diagrams, such as (b), (c) and (d) in Fig.~\ref{diagrams}, is to exploit the random-phase approximation (RPA) [see Eq. (\ref{phirpa})] which yields 
\begin{align}
\bar{\Phi}^{\rm RPA}(g_\lambda) 
= 
\frac{1}{4 \pi} \int d \omega \ln \det D_\lambda(\omega)^{-1}. 
\label{cgfph}
\end{align}
This approximation accounts for the relaxation of the phonon mode by the particle-hole excitations in the electrodes. The RPA is expected to provide results accurate 
up to second order in $\gamma$, 
\begin{align}
\bar{\Phi}^{\rm RPA}(g_\lambda) 
= 
\bar{\Phi}^{\rm (2)}(\lambda) 
+
{\mathcal O}(g^2)
\, . 
\end{align}

The dressed phonon Green function $D_\lambda$ in Eq. (\ref{cgfph}) is given by [see Appendix \ref{PHPR}] 
\begin{widetext}
\begin{align}
D_{\lambda}(\omega )^{-1}
&=
\left [
\begin{array}{cc}
(\omega_{} ^{2}-\omega^{2}_{0})/2\omega^{}_{0}-\Pi^{++}_{\lambda}(\omega )&\Pi^{+-}_{\lambda}(\omega ) \\
\Pi^{-+}_{\lambda}(\omega )&  -(\omega_{} ^{2}-\omega^{2}_{0})/2\omega^{}_{0}-\Pi^{--}_{\lambda}(\omega )
\end{array}
\right ]\ .
\label{fullphononGF}
\end{align}
\end{widetext}
In Eq. (\ref{fullphononGF}) there appears the Keldysh particle-hole propagator $\Pi$, whose lesser/greater components are given by 
\begin{align}
i \Pi^{\pm \mp}_{\lambda}(\omega)
&= 
\frac{\gamma^2}{2 \pi}  \int d \omega^{\prime}
g^{\pm \mp}_{\lambda}(\omega^\prime + \omega/2)
g^{\mp \pm}_{\lambda}(\omega^\prime - \omega/2) \  .
\label{PIo}
\end{align}
It is convenient to express these propagators in the forms
\begin{align}
i \Pi^{\pm \mp}_{\lambda}(\omega)
&= 
\sum_{{\rm r,r}^\prime={\rm L,R}}
i 
\tilde{\Pi}_{{\rm rr}^\prime \, \lambda}^{\pm \mp}(\omega)
{\rm e}^{\pm i(\lambda_{\rm r}-\lambda_{{\rm r}^\prime})}\ 
, 
\label{PI}
\end{align}
where $\tilde{\Pi}_{{\rm rr}^\prime \, \lambda}^{\pm \mp}$ describe the  particle-hole excitations   created in  the $r$ and $r'$ leads. 
Explicitly, 
\begin{align}
&{i \tilde{\Pi}_{r r^\prime \, \lambda}^{-+}(\omega)}
=
\frac{
{g {\alpha}_{r r^\prime} \rho_0^2}
}{\beta}
{\rm e}^{\beta (\omega+s_{r^\prime} V)/2}
\sum_{i}
(z_i+{\rm e}^{-\beta (s_{r} V+\omega)/2})
\nonumber \\
&\times
(z_i+{\rm e}^{-\beta (s_{r^\prime} V-\omega)/2})
\ln z_i
\prod_{j \neq i}
\frac{1}{z_i-z_j}
\ ,
\label{tildepi}
\end{align}
where 
$s_r=\pm1$ for $r=L/R$
and 
$\alpha_{rr^\prime} = {4 \Gamma_r \Gamma_{r^\prime}}/{\Gamma^2}$. 
Here $z_i$ ($i=1, \cdots, 4$) are
$z_1 = {\rm e}^{ \beta \omega/2} Z_{\lambda +}$,
$z_2 = {\rm e}^{ \beta \omega/2} Z_{\lambda -}$,
$z_3 = {\rm e}^{-\beta \omega/2} Z_{\lambda+}$, 
and 
$z_4 = {\rm e}^{-\beta \omega/2} Z_{\lambda-}$, 
where 
\begin{align}
Z_{\lambda \pm}
=&
X_\lambda
\pm 
\sqrt{
X_\lambda^2
-
1
}\ . \label{ZPM}
\end{align}
The lesser components are  obtained from the symmetry relations
\begin{align}
\Pi_\lambda^{+-}(\omega)
&=
\Pi_\lambda^{-+}(-\omega)\ 
, \nonumber
\\
\tilde{\Pi}_\lambda^{+-}(\omega)
&=
\tilde{\Pi}_\lambda^{-+}(-\omega)
\ . 
\label{sympit}
\end{align}
For $\lambda= 0$, 
Eq.~(\ref{tildepi}) reduces to the well-known form 
\begin{align}
{i \tilde{\Pi}_{r r^\prime \, 0}^{-+}(\omega)}
& =
g \alpha_{rr'} \rho_0
\frac{\mu_r-\mu_{r'}-\omega}
{{\rm e}^{\beta (\mu_r-\mu_{r'}-\omega) }-1}
\, .
\end{align}

Having determined the dressed phonon Green function Eq. (\ref{fullphononGF}), we now use it to obtain $\bar{\Phi}^{\rm RPA}$, Eq. (\ref{cgfph}). 
By using Eqs. (\ref{SPI}) and (\ref{API}), the determinant of the dressed phonon Green function is found to be
\begin{align}
-\det D_\lambda^{-1}(\omega)
=&
\left( 
\frac{\omega^2 - \omega_{0}^2}{2 \omega_0}
-{\rm Re} \Pi^R(\omega)
\right)^2
+
{\mathcal A}_\lambda(\omega)
\, , 
\label{detd}
\end{align}
where the retarded component $\Pi^R$ is expressed by the lesser and greater components upon exploiting the Kramers-Kronig relation, Eq.~(\ref{piret}). 
The counting-field dependent phonon life time broadening,  $\sqrt{{\mathcal A}_\lambda}$ is a crucial ingredient  in our analysis. It  is given by 
\begin{align}
{\mathcal A}_\lambda(\omega)
&=
\Pi^{-+}_{\lambda}(\omega) \Pi^{+-}_{\lambda}(\omega)\nonumber\\
&-\frac{1}{4}
[ \tilde{\Pi}^{-+}_{\lambda}(\omega) +  \tilde{\Pi}^{+-}_{\lambda}(\omega) 
+
\phi^S_\lambda(\omega)]^{2}
\, , 
\label{A}
\end{align}
where
$\tilde{\Pi}^{\pm \mp}_{\lambda}(\omega)=\sum_{\rm r r'} \tilde{\Pi}^{\pm \mp}_{{\rm rr'} \lambda}(\omega)$ and  
\begin{align}
i \phi^{S}_{\lambda}(\omega )
=& 
2 g \rho_0[1-2 \rho_0] 
\sum_{i}
\prod_{s,s'=\pm}
({\rm e}^{\beta (s V+s' \omega)/2} +z_i)
\nonumber \\
\times
&
\frac{
\, \ln z_i
}
{\beta z_i}
\prod_{j \neq i}
\frac{1}{z_j-z_i}\ 
, 
\label{PHILASS}
\end{align}
for the condition (\ref{wbl1}) and zero for the condition (\ref{wbl2}).

Collecting the results, we find that the derivative of Eq~(\ref{cgfph}) is
\begin{align}
\frac{\partial \bar{\Phi}^{\rm RPA}(g_\lambda)}
{\partial \lambda}
& =
\int \frac{d \omega}{4\pi}
\frac{
\partial_\lambda 
{\mathcal A}_\lambda(\omega)
}
{
\left( 
\frac{\omega^2 - \omega_0^2}{2 \omega_0} 
-
{\rm Re} \Pi^R(\omega)
\right)^2
+
{\mathcal A}_\lambda(\omega)
}
\, .
\label{phrpa}
\end{align}
The $\omega-$integrations is carried out to
leading order in $g$. 
The integrand  has four poles, which within that accuracy are given by
\begin{align}
\pm 
[\omega_0 + {\rm Re} \Pi^R(\omega_0)]
\pm i \sqrt{{\mathcal A}_\lambda(\omega_0)}
 \, , 
\label{4poles}
\end{align}
where we have  used 
${\mathcal A}_\lambda(\omega)={\mathcal A}_\lambda(-\omega)$. 
The real part of the retarded component in Eq. (\ref{4poles}) also shifts the argument of ${\mathcal A}$, which can be neglected since it is already proportional to the small parameter $g$, Eq.~(\ref{PARAG}).

The locations of poles of the integrand in Eq. (\ref{phrpa}) are complex functions of $\lambda$. Consider them first in the limit $\lambda\rightarrow 0$.
Since 
$
{\mathcal A}_0(\omega_0)
=
(
2 g 
\rho_0^2
\omega_0
)^2
$
is a positive real number, then at small enough $\lambda$, 
the real part of the squire root satisfies
${\rm Re} \sqrt{ {\mathcal A}_\lambda(\omega_0) }>0$
 and thus the integration yields 
\begin{align}
\frac{\partial \bar{\Phi}^{\rm RPA}(g_\lambda)}
{\partial \lambda}
& \approx
\partial_\lambda 
\sqrt{
{\mathcal A}_\lambda(\omega_0)
}
 \, , 
\end{align}
up to ${\mathcal O}(g)$. 
It follows that 
\begin{align}
\bar{\Phi}^{(2)}(\lambda)
& 
=
\sqrt{{\mathcal A}_\lambda(\omega_0)}
-
2 g 
\rho_0^2
\omega_0
\, , 
\label{fph}
\end{align}
which satisfies the normalization condition 
$\bar{\Phi}^{(2)}(0)=0$. 
Away from the origin $\lambda=0$,  
${\rm Re} \sqrt{ {\mathcal A}_\lambda(\omega_0) }$ may be negative. When this is the case we obtain
\begin{align}
\bar{\Phi}^{(2)}(\lambda)
& =
-
\sqrt{{\mathcal A}_\lambda(\omega_0)}
-
2 g 
\rho_0^2
\omega_0
\, . 
\label{fphm}
\end{align}
Hence a branch cut appears at $\lambda$ satisfying ${\rm Re} \sqrt{ {\mathcal A}_\lambda(\omega_0) }=0$.

In order to carry out a large-deviation analysis it suffices to consider the imaginary axis of the $\lambda-$plane, since there the rate function (\ref{ratefun}) is real. 
On the imaginary axis, ${\mathcal A}_\lambda(\omega_0)$ is a real function. 
When ${\mathcal A}_\lambda(\omega_0) > 0$
the imaginary part of $\bar{\Phi}^{(2)}$ is zero and thus the imaginary axis can serve as the steepest contour (the steepest ascent path)~\cite{BO} of the integral Eq.~(\ref{ifotr}). 
When ${\mathcal A}_\lambda(\omega_0) \leq 0$, there appears a branch cut on the imaginary axis. 
One might have thought that the branch cut would be detrimental to the saddle-point approximation of the integral in Eq.~(\ref{ifotr}). 
Here we point out that a complex integration along a path encircling the branch cut oscillates rapidly in the $\tau \to \infty$ limit and thus would be averaged out. 
Therefore we will neglect the contribution from the branch cut.

Equations  (\ref{A}), (\ref{fph}), and (\ref{fphm}), supplemented by  Eqs.~(\ref{tildepi}) and (\ref{PHILASS}),  are the main results of this paper. 
Closed expressions, but confined to the first two cumulants, have been derived in Ref. \onlinecite{Novotny}. 
In contrast, we obtain analytic expressions for the entire CGF, which enable us to examine its singularities and to fully analyze the rate function itself, as will be detailed in the Sec. \ref{RESU}.

To conclude this section we verify that our results obey the FT. 
Exploiting the extended detailed-balance relation, 
\begin{align}
\tilde{\Pi}^{-+}_{r r' \, -\lambda+i \beta V}
(\omega)
=
\tilde{\Pi}^{-+}_{r r' \, \lambda}
(\omega)
=
\tilde{\Pi}^{+-}_{r r' \, \lambda}
(\omega)
{\rm e}^{\beta(\omega-\mu_r+\mu_{r'})}
\, ,
\end{align}
and the relations 
\begin{align}
{\Pi}^{-+}_{-\lambda+i \beta V}
(\omega)
&=
{\Pi}^{+-}_{\lambda}
(\omega)
{\rm e}^{\beta \omega}
\, , \nonumber
\\
\tilde{\Pi}^{-+}_{-\lambda+i \beta V}
(\omega)
&=
\tilde{\Pi}^{+-}_{\lambda}
(\omega)
\, , \nonumber
\\
\phi^S_{-\lambda+i \beta V}
(\omega)
&=
\phi^S_{\lambda}
(\omega)
\, , 
\label{elftrel}
\end{align}
we find 
\begin{align}
{\mathcal A}_{-\lambda+i \beta V}(\omega)
=
{\mathcal A}_{\lambda}(\omega)
\, .
\end{align}
Therefore Eqs.~(\ref{fph}) and (\ref{fphm}) satisfy the FT, 
\begin{align}
\bar{\Phi}^{(2)}(-\lambda+i \beta V)
=
\bar{\Phi}^{(2)}(\lambda)
\, .
\end{align}

\section{Results and Discussion}
\label{RESU}

In the following we confine ourselves to a symmetric junction, $\alpha=1$ ($\Gamma_{\rm L}=\Gamma_{\rm R}$), and thus the normalized density of states on the dot at the Fermi level dominates the transmission probability,  ${\mathcal T}=\rho_0$ [see Eq. (\ref{T_})]. 
The numerical results are all obtained for the electron-phonon coupling constant, Eq.~(\ref{PARAG}),~$g=0.1$, unless otherwise specified.

\subsection{Average current and noise}

\begin{figure}[ht]
\includegraphics[width=.75 \columnwidth]{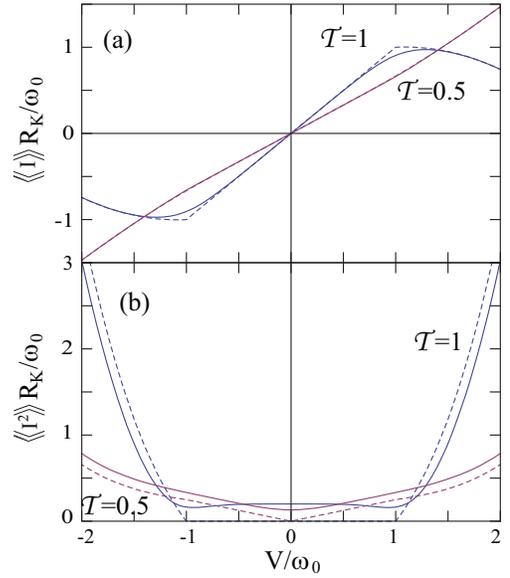}
\caption{
The source-drain bias voltage dependence of the current (a) and the current noise (b) for ${\mathcal T}=1$ and $0.5$. 
Solid lines--$\beta \omega_0=10$; 
dashed lines--zero temperature. 
The vertical axes are normalized by 
$\omega_0/R_{\rm K}$, where $R_{\rm K}=2 \pi$ is the resistance quantum. 
}
\label{is}
\end{figure}

Figure~\ref{is} (a) shows the source-drain bias voltage dependence of the current $\langle\!\langle I \rangle\!\rangle$ at a finite temperature $\beta \omega_0=10$ (solid lines) and  at zero temperature (dashed lines) for a perfect, ${\mathcal T}=1$,  and a relatively weak, ${\mathcal T}=0.5$, transmission probabilities. 
At perfect transmission the current is suppressed above the threshold $|V|>\omega_0$, because electrons are inelastically backscattered by phonons. 
When the transmission is weak, ${\mathcal T}=0.5$, the current is slightly enhanced above the threshold. 
These results are consistent with previous ones.~\cite{Haupt1,Haupt2} 
A finite temperature tends to smear the kink structure of the ${\mathcal T}=1$ curve; it affects far less the average current at weak transmission, ${\mathcal T}=0.5$, where the  solid and dashed lines almost overlap. 

Figure~\ref{is} (b) depicts the current noise $\langle\!\langle I^2 \rangle\!\rangle$. 
At perfect transmission the noise is absent below the threshold $|V|<\omega_0$ at zero temperature. 
Thermal fluctuations which arise at finite temperatures induce additional noise below the threshold. 
Although the current is suppressed above the threshold $|V|=\omega_0$ by the inelastic phonon scattering, the noise is significantly enhanced. 
This indicates that inelastic phonon scattering broadens the probability distribution of the current. In the case of a weak transmission, ${\mathcal T}=0.5$, the temperature effect is less dramatic--the noise is simply enhanced. 

\subsection{Singularities and the rate function}
\label{sandr}

At zero temperature, it is possible to obtain a simpler form for the scaled CGF, which is useful for finding its singularities. 
For positive voltages $V \geq 0$, the explicit form of the electronic part of the  CGF, Eq.  (\ref{ZO}),  reads
\begin{align}
{\cal F}^{}_{0}(\lambda )
&=
\frac{V}{2 \pi}
\ln 
\tilde{u}
\, , 
\;\;\;\;
\tilde{u}
=
1+ {\mathcal T}(u-1)
\ , 
\end{align}
where $u$ is defined in Eq.~(\ref{cft}). 
The function ${\mathcal A}_\lambda$ which determines the phonon part is given by 
\begin{align}
{\mathcal A}_\lambda(\omega_0)
=&
4 \, g^2 {\mathcal T}^2 
[
(1-1/\tilde{u})
({\mathcal T}-1)
 V 
+
{\mathcal T} \, \omega_0 
]^2
\, , 
\label{alb}
\end{align}
for $0<V<\omega_0$ and 
\begin{align}
{\mathcal A}_\lambda(\omega_0)
=&
g^2 
{\mathcal T}^2
\{ 
[ V (1-{\mathcal T} + \tilde{u}^2 (2 {\mathcal T}-1) ) + ({\mathcal T}-1 
\nonumber \\ &
+ \tilde{u} (2-\tilde{u}+2 {\mathcal T} (\tilde{u}-1) )) \, \omega_0]^2 - u {\mathcal T}^2 (V
\nonumber \\ &
-\omega_0)[2 \tilde{u} \, \omega_0+u (V+(2 \tilde{u}-1) \, \omega_0)]
\}
/ \tilde{u}^4
\, , 
\label{ala}
\end{align}
for $V \geq \omega_0$. 
Below, we investigate the analytic properties of the phonon-induced part of the CGF.

\subsubsection{Elastic phonon scattering}
\label{Elphsc}

\begin{figure}[ht]
\includegraphics[width=.85 \columnwidth]{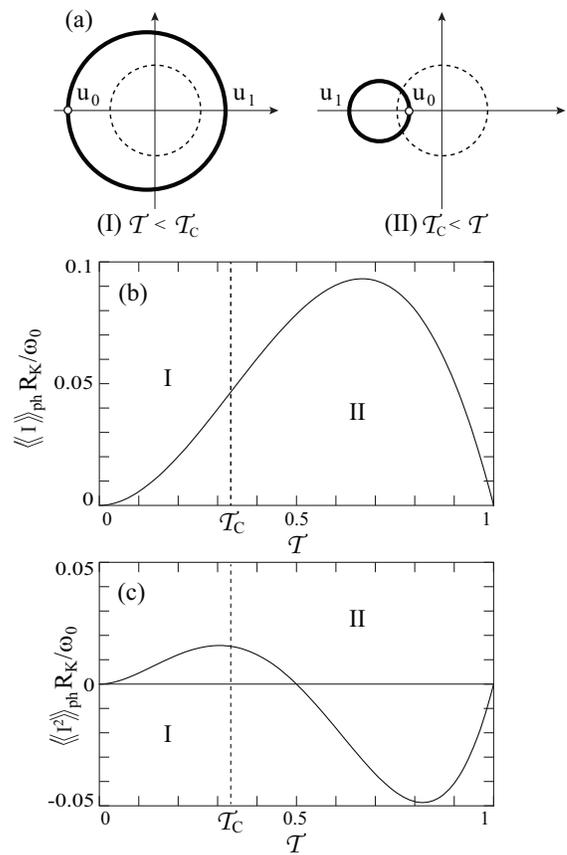}
\caption{
(a) Schematic picture of the branch cut of $\bar{\Phi}^{(2)}$ in the complex $u-$plane below  threshold, $0<V<\omega_0$. 
 The dashed circles are the unit ones. 
The corrections induced by elastic electron-phonon scattering to the average current and the current noise are portrayed in panel (b) and (c), respectively. 
The bias voltage is $V/\omega_0=0.5$. See text for the significance of ${\cal T}_{\rm C}$.
}
\label{branchb}
\end{figure}

As at zero temperature phonons cannot be excited when $0<V<\omega_0$, and electron transport at such voltages is hence affected only by {\em elastic} phonon scattering. 
Figure~\ref{branchb} (a) shows schematically the square-root branch cut of $\bar{\Phi}^{(2)}$, 
Eqs.~(\ref{fph}) and (\ref{fphm}). 
We find that the $u-$plane is separated into two by the the brach cut. 
This branch cut intersects the real axis at $u_0=1-1/{\mathcal T}$ [$u_0$ is indicated by empty dots in Fig.~\ref{branchb} (a)] and 
$u_1 = {({\mathcal T}-1) (V+\omega_0)}/[V ({\mathcal T}-1)+ {\mathcal T} \omega_0]$. 
Around these points, ${\mathcal A}_\lambda$ can be expanded as
\begin{align}
{\mathcal A}_\lambda
& \approx 
g^2 
\frac{4 V^2({\mathcal T}-1)^2}{(u-u_0)^2}
\, ,
\end{align}
and
\begin{align}
{\mathcal A}_\lambda
& \approx
g^2 
\frac{4 {\mathcal T}^4 ( {\mathcal T} \omega_0 + ({\mathcal T}-1)V)^4}
{V^2 ({\mathcal T}-1)^2}
(u-u_1)^2
\, . 
\end{align}
Upon sweeping the transmission ${\mathcal T}$ from 0 to 1, 
$u_0$ increases from $-\infty$ to $0$,
while $u_1$ increases but at ${\mathcal T}_{\rm C}=V/(V+\omega_0)$ jumps from $+\infty$ to $-\infty$. 
Accordingly, we may define two regimes, or phases, ~\cite{DIvanov}
I (${\mathcal T}<{\mathcal T}_{\rm C}$) 
and 
II (${\mathcal T}>{\mathcal T}_{\rm C}$) as indicated in Fig.~\ref{branchb} (a). 
This classification roughly captures the behavior of the average current and the current noise. 
Figures~\ref{branchb} (b) and (c) depict the corrections induced by the electron-phonon interaction in the average current and in the current noise, respectively, 
\begin{align}
\langle\!\langle I^n \rangle\!\rangle_{\rm ph}
=
\frac{\partial^n \bar{\Phi}^{(2)}}
{\partial (i\lambda )^n }
\Big |^{}_{\lambda =0}\ .
\end{align}
We find that the electron-phonon interaction always increases the average current under the conditions adopted here [Fig.~\ref{branchb} (b)]. 
On the other hand, the noise can be either enhanced or suppressed, depending on which regime the transmission is in [Fig.~\ref{branchb} (c)]. 

\begin{figure}[ht]
\includegraphics[width=.75 \columnwidth]{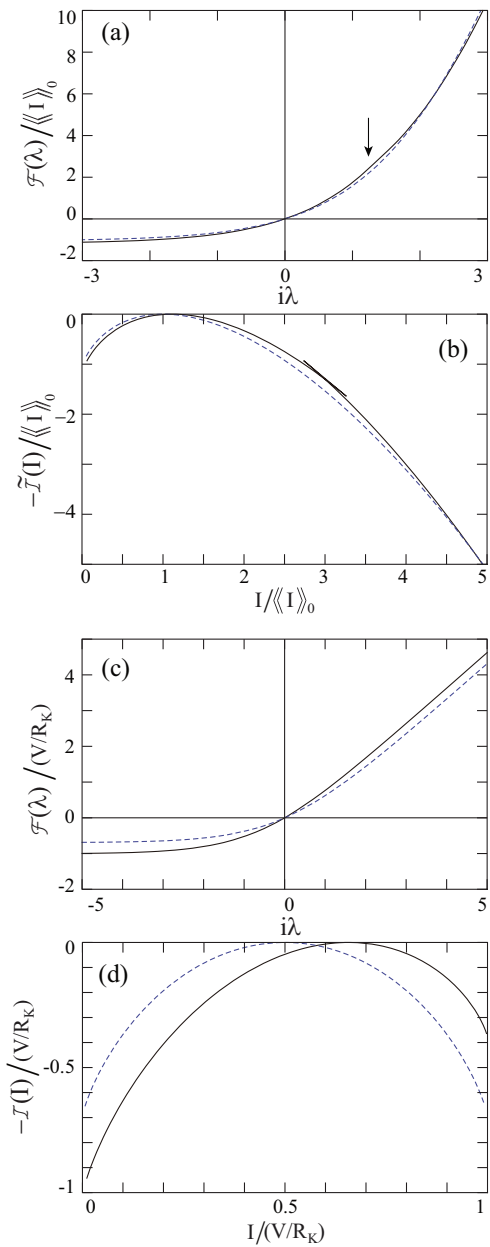}
\caption{
The scaled CGF (a) and its Legendre transform (b) in regime I (${\mathcal T}=0.1$). 
Solid lines are for $g=0.1$ and dashed lines are for $g=0$. 
The bias voltage is below threshold, $V/\omega_0=0.5$. 
The CGF is plotted as a function of the counting field on the imaginary axis. 
The axes are normalized by 
$\langle \! \langle I \rangle \! \rangle_0={\mathcal T} V/R_{\rm K}$. 
The scaled CGF (c) and the rate function (d) in regime II (${\mathcal T}=0.5$).
}
\label{cgfrateb}
\end{figure}

In regime I, one of the intersection points, $u_1$, is on the positive real axis, outside the unit circle $|u|=1$ [the left panel of Fig.~\ref{branchb} (a)]. 
Therefore, in the $\lambda-$plane, there is a nonanalytic point on the positive $i \lambda-$axis, which induces a weak non-convexity of the CGF as shown in Fig.~\ref{cgfrateb} (a). 
[The non-convex region is indicated by an arrow there. 
For comparison, we also plot the $g=0$ case (the dotted line) for which the CGF is convex.] 
Figure~\ref{cgfrateb}~(b) exhibits the Legendre transform of the CGF, 
$\tilde{ {\mathcal I} } = i \lambda^* I - {\mathcal F}(\lambda^*)$,
where $\lambda^*$ satisfies 
$I=\partial {\mathcal F}(\lambda^*)/\partial (i \lambda^*)$. 
The Legendre transform is multi-valued around $I/\langle \! \langle I \rangle \! \rangle_0 \sim 3$ (the thick line in the figure) because of the non-convexity of the CGF. 
In contrast, the Legendre-Fenchel transform,  Eq. (\ref{ratefun}), chooses the minimum value among them and provides the physical rate function ${\mathcal I}$.
[For the relation between the Legendre transform and the Legendre-Fenchel transform, see Ref.~\onlinecite{Touchette}.]  
Then, similar to the way a first-order phase transition manifests itself in thermodynamics, a kink appears in the rate function. 
We note that the location of the kink does not coincide with that of the peak; 
the peak of the rate function is at $I=\langle \! \langle I \rangle \! \rangle$ where $\lambda=0$ and ${\mathcal I}=0$. 
The physical consequence is that the elastic phonon scattering broadens the distribution by enhancing the probability of currents larger than the average value.
 It is important to note that the
kink is a consequence of the non-convexity of the CGF on the positive real axis of the $u-$plane, a feature which is absent in the noninteracting-electrons case.

In regime II the two intersection points are on the negative real axis of the $u-$plane [the right panel of Fig.~\ref{branchb} (a)]. 
The  CGF and the rate function pertaining to this case  are plotted in Figs.~\ref{cgfrateb} (c) and (d). 
The CGF is convex and the corresponding rate function is concave. 
In this regime the peak position is shifted from that for $g=0$ [dashed line]. 
The elastic scattering by the phonons can either broaden or shrink the width of the rate function depending on the transmission probability as we deduce from Fig.~\ref{branchb} (c).  

\subsubsection{Inelastic phonon scattering}
\label{Inphsc}

\begin{figure}[ht]
\includegraphics[width=.95 \columnwidth]{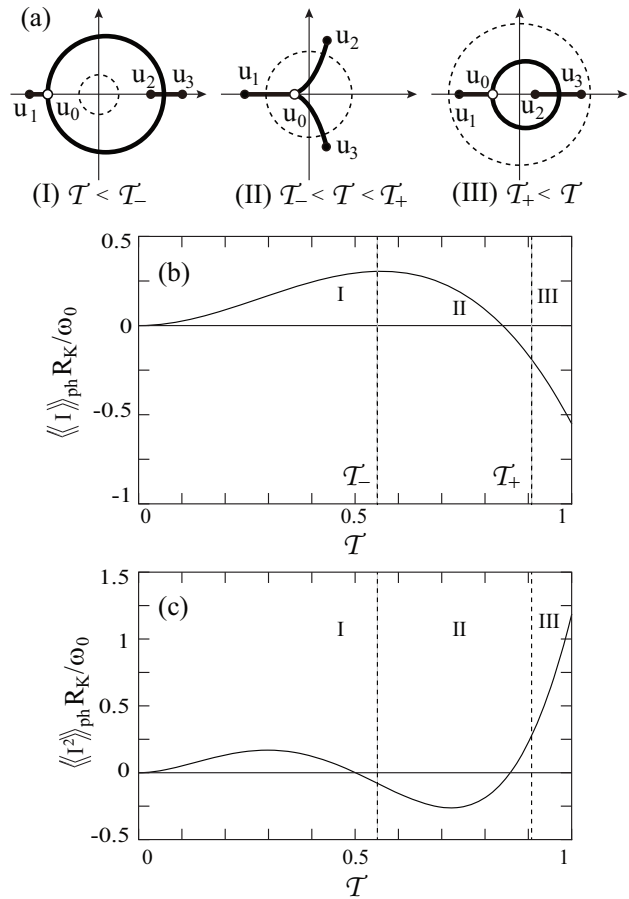}
\caption{
(a) Schematic picture of the branch cut of $\bar{\Phi}^{(2)}$ in the complex $u-$plane above threshold $V>\omega_0$. 
The dashed circles are the unit ones. 
(b) The corrections induced by the electron-phonon coupling to the current and (c) to the current noise. 
The bias voltage is $V/\omega_0=1.5$. 
The boundaries between I and II and between II and III are at 
${\mathcal T}_- \approx 0.551$ and ${\mathcal T}_+ \approx 0.908$, 
respectively. 
Regime II roughly corresponds to the negative-noise correction region. 
}
\label{brancha}
\end{figure}

Above threshold $V \geq \omega_0$,  inelastic phonon scattering becomes possible. 
The analytic properties of the scaled CGF in this regime depend on the bias voltage. 
The branch cuts in the $u-$plane are schematically shown in Fig.~\ref{brancha} (a). 
Three branch points,  $u_1$, $u_2$,  and $u_3$ can be obtained by searching for the roots of ${\mathcal A}_\lambda(u)=0$ 
[the filled dots in Fig.~\ref{brancha} (a)]. 
In addition there is another point, $u_{0}=1-1/{\mathcal T}$ 
[the empty dots in Fig.~\ref{brancha} (a)], 
around which ${\mathcal A}_\lambda$ can be expanded 
\begin{align}
{\mathcal A}_\lambda
& \approx 
g^2 
\frac{2 V({\mathcal T}-1) (\omega_0-V)}{{\mathcal T} (u-u_0)^3}
\ . 
\end{align}
The branch point $u_1$ is always on the negative real axis such that $u_1 \leq u_0$. 
From the positions of the other two branch points, $u_2$ and $u_3$, we identify three regimes, see Fig.~\ref{brancha} (a). 
In regime I the two branch points are on the positive real axis outside  the unit circle. 
As ${\mathcal T}$ increases $u_2$ and $u_3$ approach one another and meet at ${\mathcal T}={\mathcal T}_-$. 
Then in regime II, the two branch points are located symmetrically off the real axis. 
Upon further  increasing ${\mathcal T}$, $u_2$ and $u_3$ move in the complex plane and at ${\mathcal T}={\mathcal T}_+$ they meet on the real axis again. 
In regime III, the two branch points are on the positive real axis inside the unit circle. 

\begin{figure}[ht]
\includegraphics[width=.75 \columnwidth]{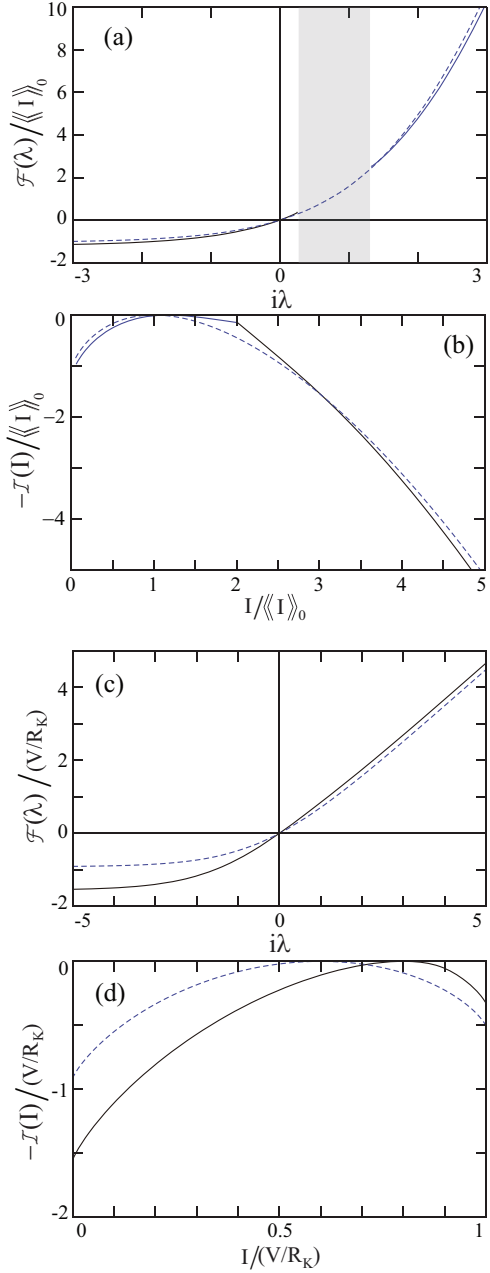}
\caption{
The CGF (a) and the rate function (b) in I (${\mathcal T}=0.1$). 
The solid lines are for $g=0.1$ and the dashed ones are for $g=0$. 
The bias voltage is above  threshold, $V/\omega_0=1.5$. 
Panels (c) and (d) show the CGF and the rate function, respectively, pertaining to II (${\mathcal T}=0.6$). 
Axes notations are the same as in Fig.~\ref{cgfrateb}.
}
\label{inel}
\end{figure}

In Figs. \ref{inel} (a) and (b) we plot the CGF and the rate function pertaining to regime I. 
The overall tendencies are similar to those found in regime I below threshold, $0<V<\omega_0$. 
In the shaded area of Fig.~\ref{inel} (a) the CGF is nonanalytic and non-convex. 
This gives rise to a stronger kink structure in the rate function [Fig.~\ref{inel} (b)]. 
Figures \ref{inel} (c) and (d) present the CGF and the rate function in regime II. 
The overall tendency is again similar to those of regime II below  threshold. 
Because of the non-analyticities off the real axis, the statistics would not be reduced to that of noninteracting electrons. 

We note that our identification of the three regimes roughly captures the behavior of the corrections to the current and the noise induced by the electron-phonon interaction [Figs.~\ref{brancha} (b) and (c)], 
which oscillate as a function of the transmission probability.~\cite{Kumar,Avriller,Schmidt}

\subsection{Fluctuation theorem}
\label{FTIII}

The analysis of the CGF and the rate function in regime III [see Fig.~\ref{brancha}] is rather subtle. 
At zero temperature, there develops a non-convex region in the CGF, and the origin $\lambda=0$ might enter it. 
When this happens, second-order perturbation fails since the rate function does not satisfy the relation ${\mathcal I}(I=\langle \! \langle I \rangle \! \rangle)=0$ [this relation is related to the normalization condition ${\mathcal F}(\lambda=0)=0$]. 
We therefore study regime III at finite temperatures, taking as an example
 $\beta \omega_0=10$, and show that the FT is crucial for obtaining a
physically-reasonable result.
Note that when the symmetry~(\ref{FFT}) holds, the FT (\ref{FT}) is also preserved within the large-deviation analysis, \cite{Lebowitz}
\begin{align}
{\mathcal I}(I)
=&
\min_\xi
\left \{
\xi I
-
{\cal F}(i \xi+i \beta V)
\right \}
\nonumber \\
=&
\min_{\xi^*}
\left \{
-
\xi^* I
-\beta I V
-{\cal F}(-i \xi^*)
\right \}
\nonumber \\
=&
{\mathcal I}(-I)
-\beta I V
\, . 
\label{FTLD}
\end{align}

\begin{figure}[ht]
\includegraphics[width=0.75 \columnwidth]{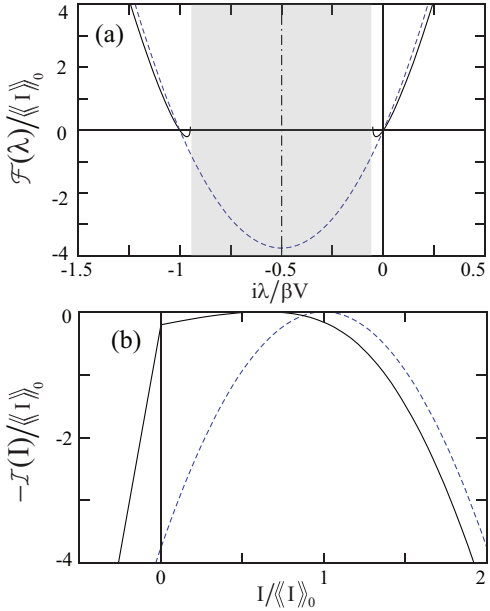}
\caption{
(a) The CGF and (b) the rate function for ${\mathcal T}=1$,  $V/\omega_0=1.5$, 
and $\beta \omega_0=10$. 
The solid (dotted) lines show results with ($g=0.1$) and without ($g=0$) electron-phonon interaction, respectively. 
In the shaded area of panel (a) the CGF for interacting case is non-analytic and non-convex, resulting in a non-differentiable point of the rate function at $I=0$ (b). 
Axes are normalized by 
$\langle\!\langle I \rangle\!\rangle_0 = V/R_{\rm K}$. 
}
\label{probdist1}
\end{figure}

Figure \ref{probdist1} exhibits the CGF and the rate function at perfect transmission. 
For comparison, we plot the corresponding curves for noninteracting electrons,  Eq. (\ref{gaussian}). 
As we have already noted when discussing the current noise, Fig.~\ref{is} (b), the width of the rate function is enhanced by inelastic phonon scattering [Fig.~\ref{probdist1} (b)]. 
The CGF obeys the FT, Eq. (\ref{FFT}), and the curves are symmetric around the  dot-dashed vertical line at $i\lambda=-\beta V/2$ [Fig.~\ref{probdist1} (a)]. 
The peak of the probability distribution is shifted in the negative direction and the probability to find large current fluctuations is suppressed as compared with the  noninteracting case [Fig.~\ref{probdist1} (b)]. 
In the shaded area of Fig.~\ref{probdist1} (a), the CGF is non-analytic and non-convex. 
Correspondingly, the rate function has a non-differentiable point at $I=0$, see  Fig.~\ref{probdist1}~(b). 
As a result, although the probability to observe  currents  smaller than the average value is enhanced by the inelastic phonon scattering, the probability to find  negative currents $I<0$ is strongly suppressed. 
This is consistent with the FT (\ref{FT}), which states that although thermal agitations generate current flowing in the opposite direction to the source-drain bias, that probability is exponentially suppressed at low temperatures. Note that previous studies report on a finite current flowing oppositely to the bias at zero temperature [see e.g. Eq. (13) in Ref.~\onlinecite{Avriller}] in disagreement with the FT, although it may be quantitatively small. 
This  can be  easily seen by calculating the probability distribution of the transmitted charge $q$  using the CGF as given by Eq. (13) of Ref.~\onlinecite{Avriller} and the inverse Fourier transform Eq.~(\ref{ifotr}), 
\begin{align}
P^{}_{\tau}(q)
&\approx
\frac{1}{2 \pi} \int_{-\pi}^\pi d \lambda 
{\rm e}^{-i \lambda q + i \bar{q}_0 \lambda + \bar{q}_1 ({\rm e}^{-i \lambda}-1)}
\nonumber \\
&=
\frac{
{\rm e}^{- \bar{q}_1} {\bar{q}_1}^{\; \bar{q}_0-q}
}
{(\bar{q}_0-q)!}
\theta(\bar{q}_0-q)
\, , 
\end{align}
{where the parameters $\bar{q}_{0}$ and $\bar{q}_{1}$ are defined in Ref.~\onlinecite{Avriller}. }
In the limit of zero temperature $\beta \to \infty$
this probability distribution remains finite at $q<\bar{q}_0$ including negative the $q$ regime, which violates the FT (\ref{FT}) .

\subsection{Discussion}

\begin{figure}[ht]
\includegraphics[width=.9 \columnwidth]{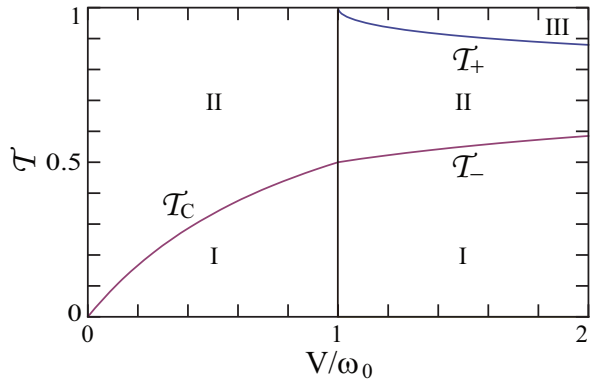}
\caption{Phase diagram summarizing the regimes discussed in Sec. \ref{sandr}. }
\label{phasediagram}
\end{figure}

Recently, Kumar {\it et al}~ \cite{Kumar} have explored experimentally  the possibility to identify  different regimes, as implied by the sign of the correction to the noise induced by the coupling with the phonons,~\cite{Schmidt,Avriller} as depicted in Fig.~\ref{brancha} (c). 
Above the threshold, our classification predicts three regimes, similarly to Ref. \onlinecite{Kumar}. 
However, the critical points quoted there, 
$
{\mathcal T}_{\pm}^{\rm Kumar} 
=
1/2 \pm 1/(2 \sqrt{2})
\approx 
0.85,0.15
$, 
are different from ours. 
Figure \ref{phasediagram} summarizes the regimes found in Sec. \ref{sandr}. 
Above the threshold, our critical points ${\mathcal T}_\pm$ depend on the bias voltage, with  ${\mathcal T}_-=1/2$ and ${\mathcal T}_+=1$ in the
$V \to \omega_0$ limit. 
Hence, there is no one-to-one correspondence between the classification of Ref.~\onlinecite{Kumar} and ours although we expect the regime II above the threshold roughly correspond to the negative phonon-induced noise regime. 
Since oscillations in higher cumulants are ubiquitous~\cite{FlindtPNAS,Flindt,Schmidt,Avriller,Golubev} and are dominated by singularities close to $\lambda=0$ as detailed in Ref.~\onlinecite{FlindtPNAS,Flindt}, 
it seems to  be legitimate to utilize the location distribution of the singularities itself for the classification.~\cite{Kambly,DIvanov,FG}

It is worthwhile to expound upon this point. 
Previous studies~\cite{Kumar,Avriller,Novotny,Schmidt} have considered the changes of slope of the $n$th cumulant at threshold, 
\begin{align}
\Delta 
\langle\!\langle I^{n} \rangle\!\rangle_{\rm ph}'
=
\left. \partial_V \langle\!\langle I^{n} \rangle\!\rangle_{\rm ph}
\right|_{V=\omega_0+0}
-
\left. \partial_V \langle\!\langle I^{n} \rangle\!\rangle_{\rm ph}
\right|_{V=\omega_0-0}
\, , 
\end{align}
as guidelines for the classifications of the various regimes. If we adopt this scheme,  we find
[see Eqs. (\ref{fph}), (\ref{alb}) and (\ref{ala})] 
\begin{align}
\Delta (\bar{\Phi}^{(2)})'
=
\left. \partial_V
\bar{\Phi}^{(2)}
\right|_{V=\omega_0+0}
-
\left. \partial_V
\bar{\Phi}^{(2)}
\right|_{V=\omega_0-0}
\ ,
\end{align}
which upon expanding becomes
\begin{align}
&\Delta (\bar{\Phi}^{(2)})'= \frac{ {\mathcal T}^2 }{2}
(5-8 {\mathcal T}) i \lambda
\nonumber \\
&+
g \frac{ {\mathcal T}^2 }{4}
(17-60 {\mathcal T}+48 {\mathcal T}^2) (i \lambda)^2
\nonumber \\
&+
g \frac{ {\mathcal T}^2 }{12}
(77-392 {\mathcal T}+642 {\mathcal T}^2-336 {\mathcal T}^3) (i \lambda)^3 
\nonumber \\
&+
\cdots
\, . 
\end{align}
We see that  the slope difference of the $n=1$ cumulant changes its sign at 
${\mathcal T}=0.635$, that of $n=2$ at ${\mathcal T}=0.434,0.816$, etc. 
In the $n=2$ case, the result is compatible with that of Ref. \onlinecite{Novotny}. 
For $n=3$, we obtain three zeros (at ${\mathcal T}=0.449,0.576,0.885$) similarly to Refs. \onlinecite{Avriller,Schmidt},  though the positions are different. In general, the coefficient of $\lambda^n$ in the expansion is an $n+2$th-degree polynomial function of ${\mathcal T}$, yielding oscillations as a function of ${\mathcal T}$, which is one example of the universal oscillations. \cite{FlindtPNAS,Berry} 
From the higher cumulants, we obtain more zeros, but it is unclear what useful information can be extracted of them. In contrast, the  topology of the singularity  distribution  is distinct and, in our opinion,  provides a better way to classify regimes where electron transport is affected differently by the coupling with the phonons. 
However,  as a direct connection between the singularity distribution  and the phonon-induced noise seems to be absent at the moment, it is not surprising that we obtain just  poor quantitative agreements between our critical points and the positions of the zeros for the phonon-induced noise in Figs.~\ref{branchb} (c) and \ref{brancha} (c).

Another comment related to previous studies has to do with the high-bias limit.   Urban {\it et. al.}~\cite{Urban} have suggested that in that limit, i.e., for  $\omega_0 \ll V \ll \Gamma$, the correction induced by the electron-phonon coupling to the cumulants  scales as
$
\langle\!\langle I^{n+1} \rangle\!\rangle_{\rm ph}/
\langle\!\langle I^n \rangle\!\rangle_{\rm ph} 
\sim 
V/\omega_0$. We can use our result Eq.~(\ref{ala}) to obtain in this regime the expansion
\begin{align}
{\mathcal A}_\lambda(\omega_0)
\approx
4 g^2 {\mathcal T}^4 \omega_0
+
2 g^2 {\mathcal T}^4 V^2 (4 {\mathcal T}-3) i \lambda
+
\cdots \,  \  ,
\end{align}
which implies a stronger scaling, 
\begin{align}
\frac{
\langle\!\langle I^{n+1} \rangle\!\rangle_{\rm ph}
}{
\langle\!\langle I^{n} \rangle\!\rangle_{\rm ph}
}
\approx
\left(n-\frac{1}{2} \right)
\left( \frac{3}{2} -2 {\mathcal T} \right)
\left( \frac{V}{\omega_0} \right)^2
\, . 
\end{align}
Our result thus  extends that of  Ref. \onlinecite{Novotny}, where $\langle\!\langle I^{2} \rangle\!\rangle_{\rm ph} \propto V^4$ was reported. We note that our phonon-induced part of the CGF, Eq. (\ref{cgfph}), is already different form that of Urban {\it et. al.}, Eq.~(11) in Ref.~\onlinecite{Urban}.

A comment on the validity of second-order perturbation theory is called for.
Flindt {\it et. al.} \cite{Flindt} have analyzed the FCS of sequential transport through a quantum dot containing two levels.~\cite{Belzig} 
They have  assumed identical incoming rates but different outgoing ones 
for the  two levels and analyzed the approximate CGF, derived by expanding in the ratio of the two outgoing rates, taken to be a small parameter. 
They have noticed a peculiar behavior: 
The agreement between the  cumulants obtained by differentiating the approximate CGF (with respect to the counting field) and those derived by differentiating the CGF before expanding it was good for the low-order cumulants,  but failed completely for the higher-order ones. 
In our case, the CGF is obtained as an expansion in the electron-phonon coupling and thus it is plausible that the unphysical results which we encountered around regime III for $V>\omega_0$ at zero temperature, may have a similar origin to the apparent discrepancy reported in Ref. \onlinecite{Flindt}. 
This fault may be resolved by accounting for all orders in the electron-phonon coupling. 
However, getting analytical results seems to be technically complicated and almost inevitably requires numerical methods. \cite{Park}

Another example of the FCS for interacting electrons  is found in transport through a quantum dot in the Kondo regime.
Recently, Sakano {\it et. al.} have calculated the CGF for the $SU(N)-$impurity Anderson model, ~\cite{Sakano} using the renormalized perturbation theory. 
Their result is exact up to  cubic order in the source-drain bias voltage ${\mathcal O}(V^3)$ for a particle-hole symmetric case,  and is quadratic in $u$. This is interpreted as the sum of the CGF for single-particle transfer and that for  two-particle transfer. 
In our case, the phonon part of the CGF at zero temperature,  to  leading order in $V$, can be obtained from Eqs.~(\ref{fph}) and (\ref{alb}) in the form 
\begin{align}
\bar{\Phi}^{(2)}
&\approx
2 g V 
\sum_{n=1}^\infty
\frac{ {\mathcal T}^{n+1} }{({\mathcal T}-1)^n}
(u^n-1)
\nonumber \\
&=
\frac{
2 g {\mathcal T} ({\mathcal T}-1)(u-1)
}{u-u_0}
V
\, , \label{SAK}
\end{align}
where $u_{0}=1-1/{\mathcal T}$.
One may interpret the first right-hand side of Eq. (\ref{SAK}) as a sum of independent $n$-electron transfers. 
However, the second line indicates that the CGF is non-analytic on the negative real axis of the $u-$plane at $u_0 \leq 0$. 
Therefore, within the first order expansion in $V$, the electron transfer statistics may be reduced to that of noninteracting electrons.~\cite{Abanov} 
This example suggests that there exist certain subtleties in interpreting the CGF for interacting electrons.

\section{summary}
\label{SUMM}

We have investigated the full-counting statistics of currents mediated by elastic and inelastic electron-phonon scattering. 
In the extended wide-band limit, we obtained analytic expressions for the cumulant generating-function, accurate up to  second order in the electron-phonon coupling. 
Our results are applicable for finite temperatures and bias voltages and satisfy the fluctuation theorem. 
Using those we analyzed the locations of singularities of the CGF. 
The singularities are symmetrically distributed in the $\lambda-$plane as to obey the fluctuation theorem. 

The singularities in $u-$ plane, which appear because of the electron-phonon interaction, classify specific regimes in which the dependence of the electron transfer statistics on the bare transmission is distinct.
For small transmission probabilities we find singularities of the CGF on the positive real axis satisfying $u>1$. 
Around the singularities, the CGF is non-convex, which results in a kink of the rate function. 
Such a kink, derived within the large-deviation analysis resembles a first-order phase transition in  thermodynamics. 
It signifies the tendency of the phonon scattering to enhance the probability to find currents larger than the average value. 

When the bias voltage is larger than the phonon frequency, $V>\omega_0$, 
we find singularities in $0<u<1$ around  perfect transmission. 
This results in a kink at $I=0$ and a strong reduction for $I<0$ in the rate function. 
This behavior can be understood in the following way: 
In this regime, phonons scatter electrons inelastically opposite to the bias. 
This broadens the rate function but 
the probability for current flowing opposite to the bias voltage is suppressed exponentially at low temperatures in accordance with the fluctuation theorem.

The kink structures in the rate function characterizes the electron-phonon interactions. 
Although measurements of the rate function of  molecular junctions would be technically demanding, the FCS can be in principle monitored experimentally, \cite{Utsumi} as is proven by existing measurements of higher cumulants~\cite{Reulet,FlindtPNAS} and of the FCS itself, ~\cite{Gustavsson,UK} for metallic and semiconducting nanostructures. This gives hope  that our predictions could be put to test.

\section*{Acknowledgments} 
We thank Christian Flindt, Dimitri Golubev, Akira Oguri and V\'aclav $\check{{\rm S}}$pi$\check{{\rm c}}$ka for valuable discussions. 
We particularly thank Tom\'{a}$\check{\rm{s}}$ Novotn\'{y} for his helpful discussion in comparing  our results with previous theories. 
This work was supported by the Bination Science Foundation (BSF)  of the US
and Israel, 
by the Israel Science Foundation (ISF),
by the Okasan-Katoh Foundation, 
by the Grant-in-Aid for Young Scientists (B) (Grants No. 23740294 and No. 24710111), 
by the Young Researcher Overseas Visits Program for Vitalizing Brain Circulation (R2214) from the JSPS, 
and 
by the MEXT kakenhi ``Quantum Cybernetics". 
This paper was written while the authors were members of a research group on molecular electronics at the Institute for Advanced Studies, Jerusalem. 


\begin{appendix}

\section{The Luttinger-Ward potential}
\label{LW}

The way to construct a partition function based on the self energy is to exploit the Luttinger-Ward functional approach, \cite{Luttinger}
or the self-consistent $\Phi$-derivable approximation. \cite{Baym}
This method can be straightforwardly extended to a nonequilibrium situation.~\cite{Ivanov} 
The saddle-point approximation for the CGF can  also be constructed by this approach. \cite{US} The underlying idea is  the observation that by introducing the Luttinger-Ward functional $\Phi$, which includes all skeleton diagrams, the total generating functional can be formally written as 
\begin{align}
{\mathcal F}(\lambda)
&=
\lim_{\tau \to \infty} \frac{1}{\tau}
\Bigl (
{\rm Tr} [\ln G^{-1}]
+{\rm Tr}[\Sigma(G) \, G]
-
\Phi(G)
\Bigr )\ ,
\label{LW1}
\end{align}
where the trace and the product mean integrations over time along the Keldysh contour. 
The Green function $G$, also defined on the Keldysh contour, is 
\begin{align}
G^{-1}(1,2)
&=
g_\lambda^{-1}(1,2)-\Sigma(1,2;G)
\  ,
\end{align}
where the arguments 1, 2, $\cdots$  stand for $t_{1}$, $t_{2}$, $\cdots$. 
Here $g_\lambda$ is the Green function of the noninteracting electrons 
\begin{align}
g_\lambda(1,2) 
&= -i \, \langle T_{\rm K }
c^{}_0(1)_{\rm I} 
c_0^{\dagger}(2)_{\rm I} \rangle
\, , 
\end{align}
given explicitly in Eqs. (\ref{GT}) and (\ref{GT1}).
The self energy is a functional of $G$ as well as of the phonon Green function $d$, 
\begin{align}
d(1,2) 
&= -i \, \langle T_{\rm K} 
(b^{}(1)_{\rm I}+b^\dagger(1)_{\rm I})
(b^{}(2)_{\rm I}+b^\dagger(2)_{\rm I}) \rangle
\  , 
\end{align}
given in Eqs. (\ref{DT}) and (\ref{DTCOM}).
The functional derivative of the nonequilibrium Luttinger-Ward functional $\Phi$ gives the self  energy, 
\begin{align}
\Sigma(1,2;G)
=
\frac{
\delta \Phi
}{
\delta G(2,1)
} 
\  . \label{LW2}
\end{align}
Both functions $G$ and $\Sigma$  depend implicitly  on the counting field only through $g_\lambda$. 

By differentiating Eq. (\ref{LW1}) with respect to the counting field $\lambda$ and using Eq. (\ref{LW2}) one obtains
\begin{align}
\frac{d {\mathcal F}(\lambda)}
{d \lambda}
=
-
\lim_{\tau \to \infty} \frac{1}{\tau}
{\rm Tr} \!
\left(
G
\frac{\partial g_\lambda^{-1}}
{\partial \lambda}
\right)
\ , 
\label{LW3}
\end{align}
where we have used the relation
\begin{align}
\frac{d \Phi}
{d \lambda}
=
{\rm Tr} \!
\left(
\frac{\delta \Phi}{\delta G}
\frac{\partial G}{\partial \lambda}
\right)
\ . 
\end{align}
It should be emphasized that in this formulation the self energy has to be determined self-consistently in order to satisfy conservation laws. \cite{Ivanov}
To second order in the electron-phonon coupling $\gamma$, Eq. (\ref{LW3}) becomes 
\begin{align}
\frac{d {\mathcal F}}{d \lambda}
& \approx
\lim_{\tau \to \infty} \frac{1}{\tau}
{\rm Tr}
\left(
g_\lambda
\frac{\partial g_\lambda^{-1}}
{\partial \lambda}
\right)\nonumber\\
&
+
\lim_{\tau \to \infty} \frac{1}{\tau}
{\rm Tr}
\left(
g_\lambda
\Sigma(g_\lambda)
g_\lambda
\frac{\partial g_\lambda^{-1}}
{\partial \lambda}
\right)
\   , \label{DIFF}
\end{align}
where we have replaced 
$\Sigma(G)$ by $\Sigma(g_\lambda)$, 
since it is  already ${\mathcal O}(\gamma^2)$. 
Upon using the identity 
\begin{align}
\frac{\partial g_\lambda}{\partial \lambda}
=
-
g_\lambda
\frac{\partial g_\lambda^{-1}}{\partial \lambda}
g_\lambda
\    , 
\end{align}
and exploiting the self-consistent condition Eq. (\ref{LW2}) [in its ${\mathcal O}(\gamma^{2})$ form], Eq. (\ref{DIFF}) becomes 
\begin{align}
\frac{d {\mathcal F}}{d \lambda}
& \approx
\lim_{\tau \to \infty} \frac{1}{\tau}
\frac{d}{d \lambda}
{\rm Tr}
[\ln 
g_\lambda^{-1}]
-
\lim_{\tau \to \infty} \frac{1}{\tau}
\frac{d \Phi(g_\lambda)}{d \lambda}
\ , 
\end{align}
and consequently 
\begin{align}
{\mathcal F}(\lambda )
& 
=
\lim_{\tau \to \infty} \frac{1}{\tau}
\Bigl (
{\rm Tr}
[\ln 
g_\lambda^{-1}]
\Bigr )
-
\bar{\Phi}^{(2)}(\lambda)
\  . 
\label{LW4}
\end{align}
Comparing this expression with the original one, Eq. (\ref{LW1}), one notes that the term depending explicitly on the self energy has disappeared. 
The lowest-order scaled Luttinger-ward potential $\bar{\Phi}^{(2)}(\lambda)$ is obtained by expanding $\bar{\Phi}(g_\lambda)$ (which may depend on $g_\lambda$ and not on $G$) up to ${\mathcal O}(\gamma^2)$ [see Eq. (\ref{slw})].

For brevity, the calculation above is presented in the time domain. 
One may also switch to the frequency representation in which Eq. (\ref{LW3}) reads 
\begin{align}
\frac{d {\mathcal F}(\lambda)}
{d \lambda}
&=&
-
\frac{1}{2 \pi}
\int d \omega
{\rm Tr} \!
\left(
G(\omega)
\tau_3
\frac{\partial g_\lambda(\omega)^{-1}}
{\partial \lambda}
\tau_3
\right) . 
\label{GK}
\end{align}
The Pauli matrix $\tau_{3}$, 
\begin{align}
{\tau_3}=
\left [
\begin{array}{cc}
1 &  0 \\
0&  -1 
\end{array}
\right ] \, . 
\label{paulimz}
\end{align}
appears once we project the time from the Keldysh contour on the real time axis, 
\begin{align}
\int_K dt
=
\int_{-\tau/2}^{\tau/2} dt_+
-
\int_{-\tau/2}^{\tau/2} dt_-
\, , \label{TINT}
\end{align}
where $t_\pm \in K_\pm$ (see Fig. \ref{Keldysh}). 
This leads to Eqs. (\ref{FFF}) and (\ref{ZORD1}) in the main text. 
Note that
Eq.  (\ref{LW3}) or (\ref{GK}) corresponds to the ``generalized current expression" given by Eq. (12) of Ref.~\onlinecite{Gogolin}. 
The latter is the starting point for several studies of FCS of molecular junctions, \cite{Haupt1,Haupt2,Novotny,Avriller} 
e.g. 
Eq. (1) in Ref. \onlinecite{Haupt1}, 
Eq. (8) in Ref. \onlinecite{Haupt2},  and
Eq. (3) in Ref. \onlinecite{Avriller}. 
We emphasize that the simple form of the generalized current expression \cite{Gogolin} is correct only when the approximate self energy of the electron Green function (\ref{LW2}) is determined self-consistently. 
This point becomes clearer when one set  the counting field in Eq. (12) of Ref.~\onlinecite{Gogolin} to zero,   $\bar{\lambda}=0$. 
That equation  is then reduced to the expression for the current,  analyzed by Hershfield, {\it et. al.}
in Refs.~\onlinecite{Hershfield}, 
where it was demonstrated that second-order perturbation for the self energy can violate  current conservation. 
Although this problem was discussed in the context of  the on-site Coulomb interaction, we suspect that it will arise for the on-site electron-phonon interaction as well.
The safe approach is to exploit the``generalized current expression",  Eq. (12) of Ref.~\onlinecite{Gogolin}, {with the self energy determined self-consistently.}

\section{Diagrammatic expansion}
\label{LCE}

Given the results of Appendix \ref{LW}, it remains to calculate $\bar{\Phi}(g_\lambda)$ up to  ${\mathcal O}(\gamma^2)$. 
This is accomplished by expanding it perturbatively in $\gamma$. 
The zeroth-order term is just a constant, 
\begin{align}
\bar{\Phi}^{(0)}(g_\lambda)
=
-
\lim_{\tau \to \infty} 
\frac{1}{2 \tau}
{\rm Tr}
\ln 
[d]
\ , \label{P0}
\end{align}
independent of the counting field. The diagrams constituting the second order are depicted in Fig.~\ref{diagrams}. 
The Hartree term [Fig.~\ref{diagrams} (a)] is 
\begin{align}
&\bar{\Phi}^{\rm H}(g_\lambda)
=
-i
\frac{\gamma^2}{2}
\lim_{\tau \to \infty} \frac{1}{\tau}
\int_{\rm K} d1 d2 \, 
g_\lambda(1,1)
d(1,2)
g_\lambda(2,2)
\nonumber 
\\
&=
-i \frac{\gamma^2}{2}
\sum_{s,s'=+,-}
s \, s'
d^{ss'}(0)
\int \frac{d \omega_1}{2 \pi} \, 
g_\lambda^{ss}(\omega_1)
\int \frac{d \omega_2}{2 \pi} \, 
g_\lambda^{s's'}(\omega_2)
\, . 
\label{DIRECT}
\end{align}
Inserting Eqs. (\ref{GT1}) and (\ref{DTCOM}) into Eq. (\ref{DIRECT}) yields
\begin{align}
\bar{\Phi}^{\rm H}(g_\lambda)
=&
\frac{g}{\pi \omega_0}
\left(
\int d \omega
\frac{1}{\Omega_\lambda(\omega)}
\frac{\Gamma (\omega-\epsilon_0)}
{(\omega-\epsilon_0)^2+\Gamma^2/4}
\right)\nonumber\\
&\times
\left(
\sum_{\rm r}
\frac{\Gamma_r}{\Gamma}
\int d \omega
\bar{\rho}(\omega)
\frac{2 f_{\rm r}^+(\omega)-1}
{\Omega_\lambda(\omega)}
\right)
\, , 
\label{hartree}
\end{align}
where the small parameter $g$ is given in Eq. (\ref{PARAG}).
Note the appearance of the distribution  Eq. (\ref{FERMI}) in the form $f_{\rm r}^+(\omega)-1/2$, resulting from the definition of the step function as $\Theta(0)=1/2$ in the continuous notation.~\cite{Kamenevbook} 
By using Eq.~(\ref{ftomega}), one can check that  the FT is fulfilled, 
$\bar{\Phi}^{\rm H}(g_{-\lambda+i \beta V})=\bar{\Phi}^{\rm H}(g_\lambda)$. 
In the main text, we neglect the Hartree term since it does not depend on the phonon distribution. 

The Fock term [Fig.~\ref{diagrams} (b)] reads
\begin{align}
\bar{\Phi}^{\rm F}(g_\lambda)
=&
i
\frac{\gamma^2}{2}
\lim_{\tau \to \infty} \frac{1}{\tau}
\int_K d1 d2 \, 
g_\lambda(1,2)
d(1,2)
g_\lambda(2,1)
\nonumber\\
=&
-\frac{1}{2}
\lim_{\tau \to \infty} \frac{1}{\tau}
\int_{\rm K} d1 d2 \, 
d(1,2)
\Pi_\lambda(2,1)
\  ,  \label{FOCK}
\end{align}
where we have introduced the particle-hole propagator, $\Pi$,  
\begin{align}
\Pi_\lambda(1,2) &= -i \gamma^2 g_\lambda(1,2) g_\lambda(2,1) \, .
\label{php}
\end{align}
Adopting the form (\ref{FOCK}), we find
\begin{widetext}
\begin{align}
\bar{\Phi}^{\rm F}(g_\lambda)
= &
-\frac{1}{2}
\int \frac{d \omega}{2 \pi} 
\biggl [
d^{++}(\omega) \Pi_\lambda^{++}(\omega) 
+
d^{--}(\omega) \Pi_\lambda^{--}(\omega) 
-
d^{-+}(\omega) \Pi_\lambda^{+-}(\omega) 
-
d^{+-}(\omega) \Pi_\lambda^{-+}(\omega) 
\biggl ]
. 
\end{align}
Here a constant should be added to keep the normalization condition 
$\bar{\Phi}^{\rm F}(g_0)=0$. 
Inserting Eqs. (\ref{SPI}) and (\ref{API}), we obtain
\begin{align}
\bar{\Phi}^{\rm F}(g_\lambda)
= &
-\frac{1}{2}
\int \frac{d \omega}{2 \pi} 
\biggl [
d^{-+}(\omega)
\biggl (
\frac{\tilde{\Pi}_\lambda^{+-}(\omega)+\Pi_0^{+-}(\omega)}{2} 
- \Pi_\lambda^{+-}(\omega)
\biggl)
+
d^{+-}(\omega)
\biggl(
\frac{\tilde{\Pi}_\lambda^{-+}(\omega)+\Pi_0^{-+}(\omega)}{2} 
- \Pi_\lambda^{-+}(\omega)
\biggl)
\nonumber \\ &
+ \frac{d^{-+}(\omega)+d^{+-}(\omega)}{2}
\phi_\lambda^S(\omega)
+ 
\frac{d^{++}(\omega)-d^{--}(\omega)}{2}
\phi_\lambda^A(\omega)
\biggl]
. 
\end{align}
Using relations (\ref{sympit}) 
and the corresponding ones for the components of the phonon Green function, e.g., 
$d^{+-}(\omega)=d^{-+}(-\omega)$, 
yields 
\begin{align}
\bar{\Phi}^{\rm F}(g_\lambda)
= &
\sum_{s=\pm}
s n^+(s \omega_0)
\biggl (
\frac{i \tilde{\Pi}_\lambda^{-+}(s \omega_0)+i {\Pi}_0^{-+}(s \omega_0)}
{2}  
-i {\Pi}_\lambda^{-+}(s \omega_0)
\biggl)
+
\frac{i}{2}
\coth
\left( \frac{\beta \omega_0}{2} \right)
\phi_\lambda^S(\omega_0)
-
{\rm P}
\int \frac{d \omega}{2 \pi}
\frac{\omega_0 \, 
\phi_\lambda^A(\omega)
}{\omega^2-\omega_0^2}
\, , 
\label{fock}
\end{align}
\end{widetext}
where ${\rm P}$ means the Cauchy principle value and $n^{\pm}$ is the Bose distribution, 
\begin{align}
n^\pm(\omega)
=\pm
\frac{ 1}
{{\rm e}^{\pm \beta \omega}-1} \, . 
\label{bosedist}
\end{align}

The Fock term depends on the equilibrium phonon distribution $n^+$, which suggests that a re-summation of an infinite series of diagrams is needed in order to account for the nonequilibrium phonon distribution. 
We carry out this summation within the random-phase approximation, by summing over all ring diagrams [see diagrams (c) and (d) in Fig. \ref{diagrams}]. 
The RPA is also known to be relevant for the AC conductance.~\cite{UEA}
In this way, we obtain the functional
\begin{align}
\bar{\Phi}^{\rm RPA}(g_\lambda)
&=
\frac{1}{2}
\lim_{\tau \to \infty} \frac{1}{\tau}
{\rm Tr} 
\ln 
\left[1- 
d \, 
\Pi_\lambda \, 
\right] 
+
\bar{\Phi}^{(0)}(g_\lambda)
\nonumber\\
&=
\frac{1}{2}
\lim_{\tau \to \infty} \frac{1}{\tau}
{\rm Tr} 
\ln 
D_\lambda^{-1}
\, , 
\label{phirpa}
\end{align}
where we have also included the zeroth-order term, Eq. (\ref{P0}) and the full phonon propagator, $D_{\lambda}$,  
\begin{align}
D_\lambda^{-1}
&=
d^{-1}-
\Pi_\lambda \, . 
\label{fpp}
\end{align}
Equation (\ref{phirpa}) yields Eq.~(\ref{cgfph}) in Fourier space. 
This RPA can be also formulated using the Keldysh path-integral approach within the saddle-point approximation~\cite{US} and further accounting for the Gaussian-fluctuation correction around it.

A comment on the   FT and  current conservation in the diagrammatic expansion is called for. 
The FT, including  current conservation as represented by Eq.~(\ref{LAM}), has been proved using  perturbation expansion in the interaction. \cite{SU} 
Although the proof has been constructed  for the  Coulomb interaction, it may be extended to the electron-phonon interaction case as well.

\section{
The electronic part}
\label{CEP}

The electronic Keldysh Green functions are obtained by inverting the matrix  Eq.~(\ref{INVG})
\begin{align}
g^{}_{\lambda}(\omega )=\left[\begin{array}{cc}g^{++}_{\lambda}(\omega) &
g^{+-}_{\lambda}(\omega)\\
g^{-+}_{\lambda}(\omega)
&
g^{--}_{\lambda}(\omega)\end{array}\right ]\ , \label{GT}
\end{align}
to obtain 
\begin{align}
&g^{ss}_{\lambda}(\omega )
=
\frac{1}{\Omega^{}_{\lambda}(\omega )}
\left (
\frac{1}{s(\omega -\epsilon^{}_{0})+i\Gamma/2}
+
is \sum_{\rm r=L,R} 
g^{s\overline{s}}_{{\rm r}}(\omega )
\right)
\ ,
\nonumber\\
&g^{s\overline{s}}_{\lambda}(\omega )
=
\sum_{\rm r=L,R}
\frac{
g^{s\overline{s}}_{{\rm r}}(\omega )
}
{\Omega^{}_{\lambda}(\omega )}
e^{is\lambda^{}_{\rm r}}
\ ,
\label{GT1}
\end{align}
where 
$\overline{s}=-/+$ for $s=+/-$. 
The lesser and greater Green functions, in the absence of the counting field, $g^{s\overline{s}}_{\rm r}$, are expressed in terms of the density of states on the localized level normalized by $\Gamma$, 
the width of the resonance on the localized level, 
\begin{align}
\bar{\rho}(\omega) &= 
\frac{\Gamma^{2}_{}/4}{(\omega-\epsilon_{0})^2+\Gamma^{2}/4}
\, , \label{DOS}
\end{align}
as
\begin{align}
g^{\pm \mp}_{\rm r}(\omega )
&=
\pm 
4 i \frac{\Gamma^{}_{\rm r}}{\Gamma^2} \bar{\rho}(\omega) 
f^{\pm}_{\rm r}(\omega )
\ . 
\end{align}
The dependence on the counting field is contained in the function $\Omega_{\lambda}$, 
\begin{align}
&\Omega^{}_{\lambda}(\omega )=-\frac{{\rm det}g^{}_{\lambda}(\omega )^{-1}_{}}{{\rm det} g^{}_{0}(\omega )^{-1}_{}}=
1+{\cal T}(\omega )\nonumber\\
&\times[f^{+}_{\rm L}(\omega )f^{-}_{\rm R}(\omega )(e^{i\lambda}-1)
+f^{+}_{\rm R}(\omega )f^{-}_{\rm L}(\omega )(e^{-i\lambda}-1)]\ ,
\label{OMEG}
\end{align}
where the transmission of the localized level is frequency dependent, 
\begin{align}
{\cal T}(\omega )
=
\alpha \bar{\rho}(\omega)
\ . \label{T}
\end{align}
From Eqs.~(\ref{ZORD1}) and (\ref{OMEG}) we can see the FT is satisfied since
\begin{align}
\Omega_{-\lambda+i \beta V}(\omega)
=
\Omega_{\lambda}(\omega)
\, . 
\label{ftomega}
\end{align}

Within the extended wide-band limit approximation, 
the frequency dependent normalized density of state~(\ref{DOS})
can be replaced by its value at the Fermi energy (\ref{APPRDOS}). 
Then the transmission becomes energy independent, as shown in Eq.~(\ref{T_}), and consequently the computation of the integral determining  the zeroth-order CGF [see  Eq. (\ref{ZORD1})] is straightforward. 
The key observation is that the variable transformation
\begin{align}
z=\exp[\beta (\omega' - (\mu_L+\mu_R)/2)] \, , 
\label{vt}
\end{align}
transforms Eq. (\ref{OMEG}) into a simpler form, 
\begin{align}
\Omega_\lambda(\omega')
=&
\frac{
(z-Z_{\lambda+})
(z-Z_{\lambda-})
}
{
(1+z \, {\rm e}^{-\beta(\mu_L-\mu_R)/2})
(1+z \, {\rm e}^{\beta(\mu_L-\mu_R)/2})
} \, , 
\end{align}
where $Z_{\lambda \pm}$ and $X_\lambda$ are given in Eq.~(\ref{ZPM}) and Eq.~(\ref{X}), respectively. 
Then we obtain 
\begin{align}
\frac{\partial {\cal F}^{}_{0}(\lambda )}{\partial (i\lambda )}
=&
\frac{-1}{2 \pi \beta}
\frac{2 \partial_{i \lambda} X_\lambda}
{Z_{\lambda+} - Z_{\lambda-}}
\ln
\frac
{Z_{\lambda-}}
{Z_{\lambda+}}
\, , 
\label{DERF0}
\end{align}
and consequently the CGF (\ref{ZO}) by integrating over $\lambda$.

\section{The phonon-induced part}
\label{PHPR}

We first derive the dressed phonon Green function $D_\lambda$ Eq.~(\ref{fullphononGF}). 
The free Keldysh phonon Green function is given by
\begin{align}
d(\omega )
=
\left[
\begin{array}{cc}
d^{++}(\omega) & d^{+-}(\omega) \\
d^{-+}(\omega) & d^{--}(\omega)
\end{array}
\right] \ , \label{DT}
\end{align}
whose four components are
\begin{align}
d^{\pm \pm}(\omega)
=&
{\rm Re}
\frac{2 \omega_0}{(\omega+i 0^+)^2-\omega_0^2}
\nonumber \\
&
-i \pi 
\coth \frac{\beta \omega}{2}
[
\delta(\omega-\omega_0)
-
\delta(\omega+\omega_0)
]
\, ,\nonumber\\
d^{\mp \pm}(\omega)
=&
-2 \pi i 
[
\delta(\omega-\omega_0)
-
\delta(\omega+\omega_0)
]
\, n^\mp(\omega)
\, . \label{DTCOM}
\end{align}
Here $0^+$ is a positive infinitesimal, and $n^{\pm}$ is the Bose distribution, Eq.~(\ref{bosedist}).  
Using the matrix form of the particle-hole Keldysh Green Function (\ref{php}), 
\begin{align}
\Pi_\lambda(\omega)
&=
\left [
\begin{array}{cc}
\Pi^{++}_{\lambda}(\omega ) & \Pi^{+-}_{\lambda}(\omega ) \\
\Pi^{-+}_{\lambda}(\omega ) & \Pi^{--}_{\lambda}(\omega )
\end{array}
\right ]\ ,
\label{mphp}
\end{align}
Eq.~(\ref{fullphononGF}) is obtained as the matrix form of Eq.~(\ref{fpp}), 
\begin{align}
D_{\lambda}(\omega )^{-1}
&=
d(\omega)^{-1}-\tau^{}_3 \Pi_\lambda(\omega) \tau^{}_3 \, ,
\end{align}
where $\tau_{3}$ is the third Pauli matrix, Eq. (\ref{paulimz}). 

Analytic expressions for the four components of the particle-hole propagator, Eq. (\ref{php}) or equivalently Eq. (\ref{PIo}), are obtained in the extended wide-band limit.~\cite{Haupt1,Haupt2,Novotny} 
Since the lesser and greater components are related to one another [see Eqs. (\ref{sympit})] it suffices to compute the greater component. 
$\tilde{\Pi}^{-+}_{{\rm rr'} \, \lambda}$, 
\begin{align}
i \tilde{\Pi}^{-+}_{{\rm rr'} \, \lambda}(\omega)
&= 
\frac{\gamma^2}{2 \pi}  \int d \omega^{\prime}
\frac{
g^{-+}_{\rm r}(\omega^\prime + \omega/2)
g^{+-}_{\rm r'}(\omega^\prime - \omega/2)
}{
\Omega^{}_{\lambda}(\omega^\prime + \omega/2)
\Omega^{}_{\lambda}(\omega^\prime - \omega/2)
}
\nonumber \\
& =
{g {\alpha}_{r r^\prime} \rho_0^2}
\int \! d \omega^{\prime}
\frac{
f_r^-(\omega_+)
f_{r^\prime}^+(\omega_-)
}
{
\Omega_{\lambda}(\omega_+)
\Omega_{\lambda}(\omega_-)
}\ ,
\end{align}
where the small parameter $g$ is given in Eq. (\ref{PARAG}), and 
\begin{align}
\omega^{}_{\pm}=\omega '\pm\omega /2 \ . 
\end{align}
After a lengthy but straightforward calculation, exploiting   the variable transformation Eq.~(\ref{vt}), we obtain Eq. (\ref{tildepi}).

The calculation of $\Pi^{++}$ and $\Pi^{--}$ is facilitated by considering the combinations $\Pi^{++}\pm\Pi^{--}$. 
The only $\lambda-$dependence of the casual and anti-casual electronic Green functions is contained in their denominator, $\Omega_{\lambda}$ 
[see  Eqs. (\ref{GT1})]. 
Therefore we may write
\begin{align}
i \Pi^{\pm \pm}_{\lambda}(\omega)
&= 
\frac{\gamma^2}{2 \pi}  \int d \omega^{\prime}
\frac{
g^{\pm \pm}_{0}(\omega_+)
g^{\pm \pm}_{0}(\omega_-)
}{
\Omega^{}_{\lambda}(\omega_+)
\Omega^{}_{\lambda}(\omega_-)
}
\, . 
\end{align}
Upon using the relations
\begin{align}
g^{++}_{0}&=g^{+-}_{0}+g_{}^{R}=g^{-+}_{0}+g_{}^{A}\ ,\nonumber\\
g^{--}_{0}&=g^{-+}_{0}-g^{R}_{}=g^{+-}_{0}-g^{A}_{}\ ,\label{IDENT}
\end{align}
where $g_{}^{R,A}$ are the retarded and advanced Green functions, 
\begin{align}
g^{R,A}_{}(\omega )=\frac{1}{\omega-\epsilon^{}_{0}\pm i\Gamma /2}\ ,
\end{align}
we find 
\begin{align}
\Pi^{++}_{\lambda}(\omega )+\Pi^{--}_{\lambda}(\omega )
&=
\tilde{\Pi}^{+-}_{\lambda}(\omega)
+
\tilde{\Pi}^{-+}_{\lambda}(\omega)
+
\phi^{S}_{\lambda}(\omega)
\ ,
\label{SPI}
\end{align}
with 
\begin{align}
i \phi^{S}_{\lambda}(\omega )
=&
\frac{\gamma^{2}}{2 \pi}\int d\omega '
\Bigl (\frac{1}{\Omega^{}_{\lambda}(\omega^{}_{+})\Omega^{}_{\lambda}(\omega^{}_{-})}-1\Bigr )\nonumber\\
& \times 
[g^{R}_{}(\omega^{}_{+})g^{R}_{}(\omega^{}_{-})+g^{A}_{}(\omega^{}_{+})g^{A}_{}(\omega^{}_{-})]
\ ,
\label{ET}
\end{align}
where the relation
\begin{align}
\int  d\omega 'g^{R/A}_{}(\omega^{}_{+})g^{R/A}_{}(\omega ^{}_{-})=0\ 
\end{align}
has been used.
Since
$\Omega_{\lambda}=1$ for $|\omega| \gg{\rm max}(|V|,1/\beta)$
the integral in Eq. (\ref{ET}) is bounded, and therefore in the
extended wide-band limit~\cite{Haupt1,Haupt2,Novotny} the terms in the square brackets there can be replaced by
\begin{align}
2 {\rm Re} \, g^{R}_{}(0)^2
\approx& \, 
\left \{
\begin{array}{cc}
8 \rho_0 [1-2 \rho_0]/\Gamma^2 & 
(|V|,k_{\rm B} T, \omega_0 \ll \Gamma)
\\
2/\epsilon_0^2 & 
(|V|,k_{\rm B} T, \omega_0, \Gamma \ll |\epsilon_0|)
\end{array}
\right. 
, 
\end{align}
yielding  Eq.~(\ref{PHILASS}); 
\begin{align}
i \phi^{S}_{\lambda}(\omega )
\approx& \, 
\left \{
\begin{array}{cc}
2 g \rho_0[1-2 \rho_0] S & 
(|V|,k_{\rm B} T, \omega_0 \ll \Gamma)
\\
0 & 
(|V|,k_{\rm B} T, \omega_0, \Gamma \ll |\epsilon_0|)
\end{array}
\right. 
, 
\end{align}
where 
\begin{align}
S&=\int d\omega '
\Bigl (\frac{1}{\Omega^{}_{\lambda}(\omega^{}_{+})\Omega^{}_{\lambda}(\omega^{}_{-})}-1\Bigr )
\nonumber \\
&=
\sum_{i}
\prod_{s,s'=\pm}
({\rm e}^{\beta (s V+s' \omega)/2} +z_i)
\frac{
\, \ln z_i
}
{\beta z_i}
\prod_{j \neq i}
\frac{1}{z_j-z_i}
. 
\end{align}

Turning now to the combination $\Pi^{++}-\Pi^{--}$, we use Eqs. (\ref{IDENT}) and the relation $g^{\rm K}=g^{-+}_{0}+g^{+-}_{0}$ for the Keldysh component of the Green function, to obtain 
\begin{align}
\Pi^{++}_{\lambda}(\omega )-\Pi^{--}_{\lambda}(\omega) 
&=
2 {\rm Re} \Pi^{R}_{}(\omega )
+
\phi^{A}_{\lambda}(\omega)
\ .
\label{API}
\end{align}
The retarded component is derived from the relation 
\begin{align}
\Pi^R(\omega )
&=
\frac{\gamma^2}{2 \pi i}
\int d \omega'
\frac{
g^{\rm K}_{}(\omega^{}_{+})g^{A}_{}(\omega^{}_{-})
+
g^{R}_{}(\omega^{}_{+}) g^{\rm K}_{}(\omega^{}_{-})
}
{2}
\ , 
\nonumber
\end{align}
which is rewritten by exploiting the Kramers-Kronig relation as
\begin{align}
\Pi^{R}_{}(\omega )
=
\frac{i}{2\pi}
\int d\omega '\frac{\Pi^{-+}_{0}(\omega ')-\Pi^{+-}_{0}(\omega ')}{\omega -\omega ' +i0^{+}}
\ . 
\label{piret}
\end{align}
In Eq.~(\ref{API}) we obtain
\begin{align}
&i \phi^{A}_{\lambda}(\omega ) = \frac{\gamma^{2}}{2\pi}\int d\omega '\Bigl (\frac{1}{\Omega^{}_{\lambda}(\omega^{}_{+})\Omega^{}_{\lambda}(\omega^{}_{-})}-1\Bigr )
\, \nonumber\\
&\times
{\rm Im} 
\Bigl(
g^{\rm K}_{}(\omega^{}_{+})g^{A}_{}(\omega^{}_{-})
+
g^{R}_{}(\omega^{}_{+}) g^{\rm K}_{}(\omega^{}_{-})
\Bigr )\ .
\end{align}
Using Eq.~(\ref{ftomega}), we can verify the symmetry 
\begin{align}
\phi^A_{-\lambda+i \beta V}(\omega)
=
\phi^A_\lambda(\omega)
\, . 
\label{ftphia}
\end{align}
One can now convince oneself that in the extended wide-band limit, in which 
${\rm Re}[g^{R}(\omega )]\approx -\epsilon_{0}/[\epsilon^{2}_{0}+\Gamma^{2}/4]$ and 
\begin{align}
g^{\rm K}(\omega )
=
-\frac{4i}{\Gamma}
\sum_{\rm r}
\frac{\Gamma^{}_{\rm r}}{\Gamma}
\rho_0
{\rm tanh}[\beta(\omega-\mu^{}_{\rm r})/2]
\label{gk}
\end{align}
is ${\mathcal O}(1/\Gamma)$, $\phi^{A}_{\lambda}$ may be safely neglected, since 
\begin{align}
\phi^{A}_{\lambda}
\propto& \, 
\left \{
\begin{array}{cc}
1/\Gamma^2 & 
(|V|, k_{\rm B} T, \omega_0 \ll \Gamma)
\\
1/(\Gamma |\epsilon_0|) & 
(|V|,k_{\rm B} T, \omega_0, \Gamma \ll |\epsilon_0|)
\end{array}
\right. 
\, . 
\label{PHYA}
\end{align}

\end{appendix}


\begin{thebibliography}{99}

\bibitem{Smit}
R. H. M. Smit, Y. Noat, C. Untiedt, N. D. Lang, M. C. van Hemert, 
and J. M. van Ruitenbeek, Nature {\bf 419}, 906 (2002). 

\bibitem{Kiguchi} 
M. Kiguchi, O. Tal, S. Wohlthat, F. Pauly, M. Krieger, D. Djukic, J. C. Cuevas, and J. M. van Ruitenbeek, Phys. Rev. Lett. {\bf 101}, 046801 (2008). 

\bibitem{Tal}
O. Tal, M. Krieger, B. Leerink, J. M. van Ruitenbeek, Phys. Rev. Lett. {\bf 100}, 196804 (2008). 

\bibitem{Agrait} 
N. Agrait, C. Untiedt, G. Rubio-Bollinger, and S. Vieira, Phys. Rev. Lett. {\bf 88}, 216803 (2002). 

\bibitem{Kumar}
M. Kumar, R. Avriller, A. Levy Yeyati, and J. M. van Ruitenbeek, 
Phys. Rev. Lett. {\bf 108}, 146602 (2012). 

\bibitem{Viljas}
J. K. Viljas, J. C. Cuevas, F. Pauly, and M. H\"afner, Phys. Rev. B {\bf 72}, 245415 (2005). 

\bibitem{delaVega}
L. de la Vega, A. Martin-Rodero, N. Agrait,  and A. Levy Yeyati, 
Phys. Rev. B {\bf 73}, 075428 (2006). 

\bibitem{Frederiksen}
T. Frederiksen, M. Brandbyge, N. Lorente, and A.-P. Jauho, Phys. Rev. Lett. {\bf 93}, 256601 (2004). 

\bibitem{Galperin}
M. Galperin, M. A. Ratner, and A. Nitzan, J. Phys. Cond. Matt. {\bf 19}, 103201 (2007). 

\bibitem{Egger} 
R. Egger and A. O. Gogolin, Phys. Rev. B {\bf 77}, 113405 (2008); O. Entin-Wohlman, Y. Imry, and A. Aharony, Phys. Rev. B {\bf 80},  035417 (2009).

\bibitem{Holstein}
T. Holstein, Ann. Phys. {\bf 8}, 325 (1959). 

\bibitem{Avriller}
R. Avriller and A. Levy Yeyati, Phys. Rev. B {\bf 80}, 041309(R) (2009).

\bibitem{Levitov}
L. S. Levitov and G. B. Lesovik, JETP Lett. {\bf 58}, 230 (1993);
L. S. Levitov, H. Lee, and G. B. Lesovik, J. Math. Phys. {\bf 37}, 4845 (1996).

\bibitem{Nazarov}
{\it Quantum Noise in Mesoscopic Physics, 
Vol. 97 of NATO Science Series II: Mathematics, Physics and Chemistry} 
edited by Yu. V. Nazarov (Kluwer Academic Publishers, Dordrecht/Boston/London, 2003).

\bibitem{Komnik}
A. Komnik and A. O. Gogolin, Phys. Rev. Lett. {\bf 94}, 216601 (2005); 

\bibitem{Gogolin}
A. O. Gogolin and A. Komnik, Phys. Rev. B {\bf 73}, 195301 (2006).

\bibitem{BUGS}
D. A. Bagrets, Y. Utsumi, D. S. Golubev, and G. Sch\"{o}n, Fortschr. Phys. {\bf 4}, 917 (2006). 

\bibitem{Bagrets}
D. A. Bagrets, 
Phys. Rev. Lett. {\bf 93}, 236803 (2004).

\bibitem{UGS}
Y. Utsumi, D. S. Golubev, and G. Sch\"{o}n, 
Phys. Rev. Lett. {\bf 96}, 086803 (2006).

\bibitem{Sakano}
R. Sakano, Y. Nishikawa, A. Oguri, A. C. Hewson, and S. Tarucha, Phys. Rev. Lett. {\bf 108}, 266401 (2012); 
R. Sakano, A. Oguri, T. Kato, and S. Tarucha, Phys. Rev. B {\bf 83}, 241301 (2011). 

\bibitem{Haupt1}
F. Haupt, T. Novotn\'{y}, and W. Belzig, Phys. Rev. Lett. {\bf 103}, 136601 (2009). 

\bibitem{Haupt2}
F. Haupt, T. Novotn\'{y}, and W. Belzig, Phys. Rev. B {\bf 82}, 165441 (2010). 

\bibitem{Novotny}
T. Novotn\'{y}, F. Haupt,  and W. Belzig, Phys. Rev. B {\bf 84}, 113107 (2011).

\bibitem{Schmidt}
T. L. Schmidt and A.  Komnik, Phys. Rev. B {\bf 80}, 041307(R) (2009).

\bibitem{Urban}
D. F. Urban, R. Avriller, and A. Levy Yeyati, Phys. Rev. B {\bf 82}, 121414(R)  (2010).

\bibitem{Simine}
L. Simine and D. Segal, 
Phys. Chem. Chem. Phys., {\bf 14}, 13820 (2012). 

\bibitem{Schaller}
G. Schaller, T. Krause, T. Brandes, and M. Esposito, arXiv:1206.3960. 

\bibitem{Maier}
S. Maier, T. L. Schmidt, and A. Komnik, Phys. Rev. B {\bf 83}, 085401 (2011). 

\bibitem{Touchette}
H. Touchette, Phys. Rep. {\bf 478}, 1 (2009). 

\bibitem{Abanov}
A. G. Abanov and D. A. Ivanov, Phys. Rev. Lett. {\bf 100}, 086602 (2008); 
A. G. Abanov and D. A. Ivanov, Phys. Rev. B {\bf 79}, 205315 (2009). 

\bibitem{Kambly}
D. Kambly, C. Flindt, and M. B\"uttiker, Phys. Rev. B {\bf 83}, 075432 (2011). 

\bibitem{DIvanov}
D. A. Ivanov, A. G. Abanov, 
Europhys. Lett. {\bf 92}, 37008 (2010). 

\bibitem{FG} 
C. Flindt and J. P. Garrahan, Phys. Rev. Lett. {\bf 110}, 050601 (2013). 

\bibitem{Yang}
C. N. Yang and T. D. Lee, Phys. Rev. {\bf 87}, 404 (1952); 
T. D. Lee and C. N. Yang, Phys. Rev. {\bf 87}, 410 (1952). 

\bibitem{Vanevic}
M. Vanevi\'c, Yu. V. Nazarov, and W. Belzig, Phys. Rev. Lett. {\bf 99}, 076601 (2007). 

\bibitem{Hassler}
F. Hassler, M. V. Suslov, G. M. Graf, M. V. Lebedev, G. B. Lesovik, 
and G. Blatter, Phys. Rev. B {\bf 78}, 165330 (2008). 

\bibitem{EG}
D. J. Evans, E. G. D. Cohen, and G. P. Morriss, Phys. Rev. Lett. {\bf 71}, 2401 (1993); {\it ibid.} {\bf 71}, 3616 (1993); 
G. Gallavotti and  E. G. D. Cohen, Phys. Rev. Lett. {\bf 74}, 2694 (1995);
G. Gallavotti, Phys. Rev. Lett. {\bf 77}, 4334 (1996).

\bibitem{Tobiska}
J. Tobiska and Yu. V. Nazarov, Phys. Rev. B {\bf 72}, 235328 (2005).

\bibitem{Foerster}
H. F\"{o}rster and M B\"{u}ttiker, Phys. Rev. Lett. {\bf 101}, 136805 (2008).

\bibitem{SU}
K. Saito and Y. Utsumi, Phys. Rev. B. {\bf 78}, 115429 (2008); arXiv:0709.4128. 
\bibitem{Andrieux}
D. Andrieux, P. Gaspard, T. Monnai, and S. Tasaki, New J. Phys. {\bf 11}
043014 (2009).

\bibitem{Esposito} 
M. Esposito, U. Harbola, and S. Mukamel, Rev. Mod. Phys. {\bf 81}, 1665 (2009).
 
\bibitem{Campisi} 
M. Campisi, P. H\"{a}nggi, and M. Talkner, Rev. Mod. Phys. {\bf 83}, 771 (2011)

\bibitem{Altland}
A. Altland, A. De Martino, R. Egger, and B. Narozhny,  
Phys. Rev. Lett. {\bf 105}, 170601 (2010); 
Phys. Rev. B {\bf 82}, 115323 (2010).  

\bibitem{Lopez}
R. Lopez, J. S. Lim and D. Sanchez, Phys. Rev. Lett. {\bf 108}, 246603 (2012); 
R. Sanchez, R. Lopez, D. Sanchez, and M. B\"{u}ttiker, Phys. Rev. Lett. {\bf 104}, 076801 (2010); 
J. S. Lim, D. Sanchez, and R. Lopez, Phys. Rev. B {\bf 81}, 155323 (2010); 
D. Sanchez, Phys. Rev. B {\bf 79}, 045305 (2009). 

\bibitem{UK}
Y. Utsumi, D. S. Golubev, M. Marthaler, K. Saito, T. Fujisawa, and G. Sch\"{o}n, Phys. Rev. B {\bf 81}, 125331 (2010);
B. K\"{u}ng, C. R\"{o}ssler, M. Beck, M. Marthaler, D. S. Golubev, Y. Utsumi, T. Ihn, and K. Ensslin, Phys. Rev. X {\bf 2}, 011001 (2012).

\bibitem{Nakamura}
S. Nakamura, Y. Yamauchi, M. Hashisaka, K. Chida, K. Kobayashi, 
T. Ono, R. Leturcq, K. Ensslin, K. Saito, Y. Utsumi, and A. C. Gossard, 
Phys. Rev. Lett. {\bf 104}, 080602 (2010); 
Phys. Rev. B {\bf 83}, 155431 (2011). 

\bibitem{US}
Y. Utsumi and K. Saito, Phys. Rev. B. {\bf 79}, 235311 (2009).

\bibitem{Saira}
O.-P. Saira, Y. Yoon, T. Tanttu, M. M\"ott\"onen, D. V. Averin, and J. P. Pekola, 
Phys. Rev. Lett. {\bf 109}, 180601 (2012).

\bibitem{Baym}
G. Baym and L. P.  Kadanoff, Phys. Rev. {\bf 124}, 287 (1961); G. Baym,
Phys. Rev. {\bf 127}, 1391 (1962). 

\bibitem{Luttinger}
J. M. Luttinger and J. C. Ward, Phys. Rev. {\bf 118}, 5 (1960); 
J. M. Luttinger, Phys. Rev. {\bf 119}, 4 (1960). 

\bibitem{Ivanov}
Yu. B. Ivanov, J. Knoll, H. Van Hees, and D. N. Voskresensky, Nucl. Phys. A {\bf 657}, 413 (1999). 

\bibitem{BO}
C. M. Bender and S. A. Orszag, 
{\it Advanced Mathematical Methods for Scientists and Engineers}, 
(Springer, New York, 1999). 

\bibitem{Mitra}
A. Mitra, I. Aleiner, and A. J. Millis, Phys. Rev. B {\bf 69}, 245302 (2004).

\bibitem{ora}
O. Entin-Wohlman, Y. Imry, and A. Aharony, 
Phys. Rev. B {\bf 81}, 113408 (2010). 

\bibitem{Rosch}
A similar problem has been encountered for a local moment coupled with a nonequilibrium electron gas, see e.g.
A. Rosch, J. Paaske, J. Kroha, and P. W\"{o}lfle, Phys. Rev. Lett. {\bf 90}, 076804 (2003); 
O. Parcollet and C. Hooley, Phys. Rev. B {\bf 66}, 085315 (2002).


\bibitem{Park}
T.-H. Park and M. Galperin, Phys. Rev. B {\bf 84}, 205450 (2011).

\bibitem{Lebowitz}
J. L. Lebowitz and H. Spohn, 
J. Stat. Phys. {\bf 95}, 333 (1999). 

\bibitem{FlindtPNAS} 
C. Flindt, C. Fricke, F. Hohls, T. Novotn\'{y}, K. Neto\`{c}n\'{y}, T. Brandes, and R. J. Haug, 
Proc. Natl. Acad. Sci. USA {\bf 106}, 10116 (2009).

\bibitem{Golubev} 
D. S. Golubev, M. Marthaler, Y. Utsumi, and Gerd Sch\"on, 
Phys. Rev. B {\bf 81}, 184516 (2010).

\bibitem{Flindt} 
C. Flindt,  T. Novotn\'{y}, A. Braggio, and A-P. Jauho, Phys. Rev. B {\bf 82}, 155407 (2010).

\bibitem{Berry}
M. V. Berry, 
Proc. R. Soc. A {\bf 461}, 1735 (2005). 

\bibitem{Belzig}
W. Belzig, Phys. Rev. B {\bf 71}, 161301(R) (2005). 

\bibitem{Utsumi}
Y. Utsumi, D. S. Golubev, M. Marthaler, G. Sch\"on, and K. Kobayashi, Phys. Rev. B {\bf 86}, 075420 (2012). 

\bibitem{Reulet}
B. Reulet, J. Senzier, and D. E. Prober, Phys. Rev. Lett. {\bf 91}, 196601 (2003); 
Yu. Bomze, G. Gershon, D. Shovkun, L. S. Levitov, and M. Reznikov, Phys. Rev. Lett. {\bf 95}, 176601 (2005).

\bibitem{Gustavsson}
S. Gustavsson, R. Leturcq, B. Simovic, R. Schleser, T. Ihn, P. Studerus, K. Ensslin, D. C. Driscoll, and A. C. Gossard, Phys. Rev. Lett. {\bf 96}, 076605 (2006); T. Fujisawa, T. Hayashi, R. Tomita, and Y. Hirayama, Science {\bf 312}, 1634 (2006).

\bibitem{Kamenevbook}
For a precise treatment of the step function, 
see Sec. 2.8 in 
A. Kamenev, 
{\it Field Theory of Nonequilibrium Systems} (Cambridge University Press, Cambridge, 2011). 

\bibitem{Hershfield}
S. Hershfield, J. H. Davies, and J. W. Wilkins, 
Phys. Rev. Lett. {\bf 67}, 3720 (1991); 
S. Hershfield, J. H. Davies, and J. W. Wilkins, 
Phys. Rev. B {\bf 46}, 7046 (1992). 

\bibitem{UEA}
A. Ueda, O. Entin-Wohlman, and  A. Aharony, 
Phys. Rev. B {\bf 83}, 155438 (2011). 



\end{thebibliography}
\end{document}